\documentclass[%
 reprint,
 superscriptaddress,
nofootinbib,
 floatfix,
 amsmath, amssymb,
 aps,
 prx,
]{revtex4-2}

\usepackage{ragged2e}
\usepackage{enumerate}
\usepackage{graphicx}
\usepackage[font=small]{caption}
\usepackage{subcaption}
\usepackage{dcolumn}
\usepackage{bm}
\usepackage[colorlinks=true,linkcolor=blue,citecolor=blue,urlcolor=blue]{hyperref}
\usepackage{amsthm}
\newtheorem*{main_res}{Main result}
\usepackage{amsmath}
\usepackage{amssymb}
\usepackage{soul}
\usepackage{dsfont}
\usepackage{comment}
\usepackage{xcolor}
\definecolor{BlueGreen}{rgb}{0.0, 0.87, 0.87}
\usepackage{braket}
\usepackage{mathtools}
\usepackage[linesnumbered,ruled,vlined]{algorithm2e}
\newtheorem{proposition}{Proposition}

\newtheorem{thm}{\protect\theoremname}
\theoremstyle{plain}

\theoremstyle{plain}

\theoremstyle{plain}
\newtheorem*{lem*}{\protect\lemmaname}
\theoremstyle{plain}

\theoremstyle{plain}

\newtheorem{defn}{Definition}
\newtheorem{example}{Example}

\usepackage{bbm}
\newcommand{\bbone}{\mathbbm{1}}

\newcommand{\matM}{\mathsf{M}}

\usepackage[USenglish]{babel}
\providecommand{\corollaryname}{Corollary}
\providecommand{\lemmaname}{Lemma}
\providecommand{\propositionname}{Proposition}
\providecommand{\remarkname}{Remark}
\providecommand{\theoremname}{Theorem}

\usepackage[normalem]{ulem}

\newcommand{\cN}{\mathcal{N}}
\newcommand{\Err}{\mathsf{E_{rr}}}

\definecolor{todoteal}{HTML}{008080}
\definecolor{citeblue}{HTML}{0055cc}
\hypersetup{colorlinks=true, 
            linkcolor=blue,
            citecolor=blue,
            urlcolor=blue
}
\begin{document}

\title{Rigorous Time-dependent Hamiltonian Learning via Continuous Weak Measurements}

\author{Jesús Jiménez-Rodríguez}
\email{jesus.jimenez.rodriguez@uab.cat}
\affiliation{ICFO-Institut de Ciencies Fotoniques, The Barcelona Institute of Science and Technology, 08860 Castelldefels (Barcelona), Spain}
\affiliation{Física Teòrica: Informació i Fenòmens Quàntics, Departament de Física, Universitat Autònoma de Barcelona, 08193 Bellaterra (Barcelona), Spain}
\author{Giacomo Franceschetto}
\affiliation{ICFO-Institut de Ciencies Fotoniques, The Barcelona Institute of Science and Technology, 08860 Castelldefels (Barcelona), Spain}
\author{Antonio Acín}
\affiliation{ICFO-Institut de Ciencies Fotoniques, The Barcelona Institute of Science and Technology, 08860 Castelldefels (Barcelona), Spain}
\affiliation{ICREA-Institució Catalana de Recerca i Estudis Avançats, Lluís Companys 23, 08010 Barcelona, Spain}
\author{Luciano Pereira}
\email{luciano.pereira@icfo.eu}
\affiliation{ICFO-Institut de Ciencies Fotoniques, The Barcelona Institute of Science and Technology, 08860 Castelldefels (Barcelona), Spain}

\date{\today}
\begin{abstract}

Characterizing the Hamiltonian that a quantum processor actually implements is central to calibrating and validating current quantum hardware. Many devices, however, operate with generators that are time-dependent by design.
Here we develop a rigorous and experimentally friendly protocol for learning time-dependent many-body Hamiltonians from continuous weak-measurement signals. The key observation is that interaction sparsity reduces the global reconstruction to a set of local inverse problems, whose size is controlled by the interaction connectivity rather than by the system size. Pure separable probe states suffice to drive these inversions, and a graph-coloring construction embeds them into a small number of global product-state preparations. We derive explicit reconstruction-error bounds and a sample-complexity theorem that cleanly separates the finite-sampling statistical noise from the deterministic bias, which arises from the derivative approximation and the state propagation, and we validate the protocol on time-dependent spin chains with up to \(n=8\) qubits. Beyond these results, our analysis provides a rigorous foundation for time-dependent Hamiltonian learning from continuous monitoring in many-body systems, establishing a framework that extends naturally to many platforms and probe ensembles.

\end{abstract}

\maketitle

\section{Introduction}

Quantum hardware has advanced rapidly, with programmable processors and analog simulators now reaching tens to hundreds of interacting degrees of freedom~\cite{Preskill2018NISQ,Georgescu2014QuantumSimulation,Ebadi2021QuantumPhases}. This progress creates a strong need for methods that characterize and calibrate not just the outputs these devices produce, but the dynamics they actually implement. Hamiltonian learning (HL) addresses this need directly at the level of the closed-system generator: rather than certifying an output state or an effective input--output map, it infers the generator of the evolution from experimental data~\cite{certification,Eisert2020QuantumCertification}, exposing the interactions responsible for deviations from the intended model, whether they originate from noise, residual couplings, or an incomplete description of the device.

Several complementary approaches have been developed to make this reconstruction task practical under different assumptions on available states, measurements, controls, and noise \cite{Granade2012,certification,PhysRevA.89.042314,anshu2024,zeno,Huang2023HeisenbergHL,Wang2015HamiltonianTomography,Haah2024,Yu2023,StilckFrana2024,Heightman2025,BaranHeisenberg,Pastori2022LiouvillianLearning,Olsacher2025HamiltonianLiouvillianLearning, PhysRevLett.122.020504, Hangleiter2024, Guo2025}. Across these settings, however, the central challenge remains scalability: in the absence of additional structure, the number of Hamiltonian parameters grows exponentially with system size. The decisive simplifying principle is locality. Physically relevant Hamiltonians are typically sparsely interacting, with each term overlapping with only a bounded number of others. Exploiting this locality turns an otherwise global reconstruction into a scalable one, and has led to rigorous protocols that learn sparse Hamiltonians from unitary dynamics or Gibbs states~\cite{anshu2024}, from reshaped dynamics that isolate local interactions~\cite{Huang2023HeisenbergHL,Wang2015HamiltonianTomography}, or through the quantum Zeno effect~\cite{zeno}.

These protocols share a key assumption: the generator is static, or effectively time-independent. Yet in many hardware-relevant settings the effective Hamiltonian is explicitly time dependent by design. Single- and two-qubit gates are driven by shaped control pulses~\cite{Chow2010OptimizedDriving}; interactions are switched and modulated through tunable couplers~\cite{Chen2014TunableCoupling,Sete2021ParametricResonance}; and programmable simulators use time-dependent controls to drive quenches, engineer Floquet phases, and realize nonequilibrium protocols~\cite{Ebadi2021QuantumPhases,Randall2021TimeCrystal}. In all these cases, the relevant object is not a single effective Hamiltonian but the full trajectory $H(t)$ followed during the evolution.

This motivates the development of HL protocols for time-dependent Hamiltonians, whose rigorous treatment has only recently been addressed. The first rigorous approach follows an evolve-and-measure strategy, in which the system is prepared in a known state, evolved up to a chosen time, and projectively measured~\cite{StilckFranca2025LocalTimeDependent}. Assuming a known ansatz for the time dependence, this strategy learns local time-dependent Hamiltonians with sample complexity that is logarithmic in the system size. Each run, however, ends with a destructive measurement and provides information only at a single time, so resolving the full trajectory $H(t)$ requires restarting the evolution at every time of interest.

Addressing this limitation calls for a probe that resolves the dynamics in time without interrupting it. Continuous weak measurements provide exactly this. By weakly monitoring selected observables while the system evolves, they extract time-resolved information while retaining part of the coherent dynamics~\cite{Jacobs2006ContinuousMeasurement}. Such monitoring is by now an established experimental capability, demonstrated in superconducting circuits~\cite{Murch2013Trajectories,Weber2016Trajectories,HacohenGourgy2020Continuous}, trapped ions~\cite{Pan2020WeakToStrongIon}, cold atomic ensembles and Bose--Einstein condensates~\cite{Smith2004ContinuousWeakMeasurement,Smith2006ContinuousWeakMeasurement,Murch2008BackactionUltracoldGas,Altuntas2023WeakMeasurementBEC}, and optomechanical resonators~\cite{Rossi2019MechanicalTrajectory}. The same effective weak-measurement dynamics can be engineered digitally through repeated ancilla-mediated weak measurements with tunable strength, mid-circuit measurement, and reset, thereby enabling the realization of the corresponding measurement-induced dephasing on cloud-accessible quantum processors \cite{franceschetto2026measurementenabledonlinequantumprocessing}. This maturity makes continuous weak measurements a natural resource for characterizing dynamics, and they were recently used to reconstruct unknown time-dependent Hamiltonians in one- and two-qubit superconducting systems~\cite{PRXQuantum.4.040324}. That demonstration, however, was limited to few-qubit systems, without guarantees on the required probes and sample complexity. Whether the locality that made static HL scalable can also enable scalable time-dependent reconstruction from weak-measurement data has, so far, remained an open question.

In this work, we establish a rigorous framework for time-dependent HL in sparse many-body systems from continuous weak measurements. Our contribution can be summarized in four points: (i)~time-dependent HL based on weak measurements extends to sparse many-body systems by reducing the global reconstruction to local inverse problems whose size is set by the interaction connectivity rather than by the system size; (ii)~pure separable probe states already suffice to initialize these local inversions, so no entangling preparations are required; (iii)~a graph-coloring construction embeds the required local probes into a small number of global product-state preparations, with an overhead controlled by the interaction degree rather than the system size; and (iv)~a rigorous error analysis cleanly separates the finite-trajectory statistical noise from the deterministic bias arising from the derivative approximation and the state propagation, yielding explicit reconstruction-error bounds and a sample-complexity theorem.

Concretely, we consider Hamiltonians with known sparse Pauli structure and unknown time-dependent coefficients. In each run, the system is prepared in a separable probe state, evolves under the unknown Hamiltonian, and is continuously monitored through local weak measurements, whose signals drive the local inversions above. Because the probes are separable and the monitoring is local, the measurement model stays close to near-term analog platforms while still admitting a fully rigorous error analysis.

These ingredients yield concrete guarantees. For sparsely interacting Hamiltonians, the number of probe configurations is set by the local interaction degree and does not grow with the system size; we give an explicit probe set from a graph coloring of the interaction structure, an $\ell_2$ reconstruction-error bound, and a sample-complexity theorem for the number of trajectories needed to reach a prescribed accuracy with high probability. We validate the protocol numerically on time-dependent spin chains with up to $n=8$ qubits, recovering the full coefficient trajectories and reproducing the predicted crossover from a sampling-limited regime to a deterministic bias floor. More broadly, these results establish a rigorous foundation for time-dependent HL from continuous monitoring in many-body systems, and we expect the underlying structure to extend to richer geometries, probe ensembles, and open-system generators.

The remainder of this article is organized as follows. Section~\ref{sec:preliminaries} introduces the HL setting, the continuous weak-measurement model, and the graph-theoretic notions used to encode sparse interactions. Section~\ref{sec:protocol} presents the reconstruction protocol, deriving the local inverse problems from the weak-measurement signals and explaining the iterative state-update procedure. Section~\ref{sec:probe_configurations} provides the graph-coloring construction of separable probe configurations. Section~\ref{sec:performance} establishes the reconstruction-error bound and the corresponding sample-complexity guarantee. Section~\ref{sec:numerics} presents numerical simulations validating the protocol on time-dependent spin models. Finally, Section~\ref{sec:conclusion} concludes with a discussion and possible future directions.

\section{Preliminaries}
\label{sec:preliminaries}

We begin by stating the learning problem addressed in this work. Our goal is to estimate the time-dependent coefficients of a quantum many-body Hamiltonian from continuous weak-measurement signals. More precisely, we consider an $n$-qubit system evolving under a time-dependent Hamiltonian of the form
\begin{equation}
    H(t)=\sum_{i=1}^{r}\mu_i(t)W_i,
    \label{eq:hamiltonian_ansatz}
\end{equation}
where each $W_i\in\{\bbone,\sigma_x,\sigma_y,\sigma_z\}^{\otimes n}$ is a Pauli string and
$\boldsymbol{\mu}(t)=(\mu_1(t),\ldots,\mu_r(t))^T\in\mathbb{R}^r$ denotes the vector of Hamiltonian coefficients. We assume that $H(t)$ is traceless, so that $W_i\neq\bbone^{\otimes n}$, and that the operator structure $\{W_i\}_{i \in [r]}$ is known. The unknowns are the real time-dependent coefficients $\mu_i(t)$, which determine the instantaneous Hamiltonian.

Information about $H(t)$ is extracted from repeated runs of the monitored dynamics. In each run, we prepare a known initial state and let it evolve under the unknown Hamiltonian while continuously and weakly measuring the local Pauli observables $\sigma_z^{(j)} \coloneq \bbone ^{\otimes j-1}  \otimes \, \sigma_z \otimes   \bbone ^{\otimes n-j}$. Each run thus yields a time-resolved signal over the entire evolution, rather than a single readout at its end. We refer to each choice of initial state as a probe configuration and repeat the run many times for each configuration. From the resulting weak-measurement signals, the task is to construct an estimator $\hat{\boldsymbol{\mu}}(t)
    =
    (\hat{\mu}_1(t),\ldots,\hat{\mu}_r(t))^T$
such that, for a fixed reconstruction time $t$ and prescribed accuracy $\epsilon>0$,
\begin{equation}
    \|\hat{\boldsymbol{\mu}}(t)-\boldsymbol{\mu}(t)\|_2\leq \epsilon,
\end{equation}
with probability at least \(1-\delta\), where
\begin{equation}
    \|\hat{\boldsymbol{\mu}}(t)-\boldsymbol{\mu}(t)\|_2
    =
    \left(\sum_{i=1}^{r}
    |\hat{\mu}_i(t)-\mu_i(t)|^2\right)^{1/2}.
\end{equation}
We now introduce the ingredients needed to formulate the protocol. First, we describe the weak continuous-measurement model and the effective backaction it induces on the system dynamics. We then introduce the interaction graph and the local neighborhoods that determine which Hamiltonian coefficients are visible in each measurement signal.

\subsection{Continuous weak measurements and backaction}
\label{subsec:weak_measurements}

Continuous weak measurements extract time-resolved information from a quantum system without interrupting its evolution by a sequence of projective measurements. In contrast with a strong measurement, which maximally reveals the measured observable while strongly disturbing the state, a weak measurement acquires only partial information at each time step. A tunable measurement strength sets the trade-off between information gain and disturbance, so that the system can be monitored continuously while retaining part of its coherent dynamics~\cite{Jacobs2006ContinuousMeasurement,WisemanMilburn2009,Hatridge2013BackAction,Mujal2023WeakProjective}.

A standard way to implement such monitoring is dispersive readout~\cite{Blais2021CircuitQED}. Each qubit is coupled to its own off-resonant cavity, so that the coupling only shifts the cavity frequency by an amount that depends on the qubit state in the computational basis. This scheme is used across several platforms, including atoms in optical cavities~\cite{Bochmann2010LosslessDetection} and spin qubits~\cite{West2019GateBasedReadout}, and is the standard readout of superconducting circuits~\cite{Blais2021CircuitQED}. When the cavity is driven, the outgoing field acquires a qubit-dependent phase, which homodyne detection converts into a continuous current~\cite{Clerk2010QuantumNoise,WisemanMilburn2009}.

If the cavities respond fast compared with the qubit dynamics, the average state of the qubits evolves according to the master equation~\cite{Gambetta2008QuantumTrajectory} (\(\hbar=1\))
\begin{equation}
    \frac{d\rho(t)}{dt}
    =
    -i[H(t),\rho(t)]
    +
    \sum_{j=1}^{n}
    \frac{\Gamma_j}{2}
     D[Z_j]\rho(t),
    \label{eq:wm_master_equation}
\end{equation}
where \(Z_j\coloneq\sigma_z^{(j)}\) for \(j\in[n]\), \(D[L]\rho=L\rho L^{\dagger}-\frac{1}{2}\{L^{\dagger}L,\rho\}\) is the Lindblad dissipator, and \(\Gamma_j\) is the measurement-induced dephasing rate, set by the readout power~\cite{Schuster2005ACStark,Gambetta2008QuantumTrajectory}. The dissipative term is the backaction of the measurement~\cite{Clerk2010QuantumNoise,Hatridge2013BackAction}. The same coupling that produces the signal dephases each qubit in the monitored basis, so the state does not evolve according to a closed Schrödinger equation. 

In a single run of the experiment, the homodyne current associated with qubit \(j\) takes the form~\cite{WisemanMilburn2009,Jacobs2006ContinuousMeasurement,Gambetta2008QuantumTrajectory}
\begin{equation}
    I_j(t)=\sqrt{2\Gamma_j}\,\langle Z_j\rangle_{\mathrm{c}}(t)+\xi_j(t),
    \label{eq:homodyne_current}
\end{equation}
where \(\langle Z_j\rangle_{\mathrm{c}}(t)\) is the expectation value of \(Z_j\) in the qubit state conditioned on the current observed up to time \(t\), and \(\xi_j(t)\) is Gaussian white noise with \(\mathbb{E}[\xi_j(t)\xi_k(t')]=\delta_{jk}\delta(t-t')\). At each instant, the current thus carries a small signal proportional to \(Z_j\), hidden in the noise, and \(\Gamma_j\) also sets how fast information about \(Z_j\) accumulates~\cite{Clerk2010QuantumNoise}. Averaging the current over many repetitions of the experiment suppresses the noise and yields the ideal measurement signal associated with site \(j\),
\begin{equation}
    z_j(t)=\mathrm{Tr}\!\left[Z_j\rho(t)\right],
    \label{eq:ideal_measurement_record}
\end{equation}
with \(\rho(t)\) the solution of Eq.~\eqref{eq:wm_master_equation}. In practice, this quantity is estimated from a finite number of noisy weak-measurement trajectories. We denote the corresponding empirical estimator by
\begin{equation}
    \hat z_j(t)
    =
    \frac{1}{N_s}
    \sum_{\nu=1}^{N_s}
    z_{j,\nu}(t),
    \label{eq:empirical_measurement_record}
\end{equation}
where \(N_s\) is the number of trajectories and \(z_{j,\nu}(t)\) is the homodyne current of Eq.~\eqref{eq:homodyne_current} in trajectory \(\nu\), rescaled by \(1/\sqrt{2\Gamma_j}\). We assume that the signals are sampled with time resolution \(\Delta t\), and write \(t_k=k\Delta t\). The data available to the reconstruction are therefore the discrete time series \(\hat z_j(t_k)\), collected for each monitored site and for each chosen probe preparation.

Increasing the measurement strength improves the signal-to-noise ratio of the homodyne signal, but it also increases the disturbance induced by the measurement and shortens the time over which coherent dynamics can be observed. The protocol therefore operates in the weak-measurement regime, where the monitoring is strong enough to produce a resolvable signal but weak enough not to dominate the dynamics during the reconstruction window~\cite{Hatridge2013BackAction,Murch2013Trajectories,Mujal2023WeakProjective}. Denoting by \(t_{\mathrm{rec}}\) the time interval over which the Hamiltonian is reconstructed, this requires
\begin{equation}
    t_{\mathrm{rec}} \ll \min_j \Gamma_j^{-1}.
    \label{eq:weak_measurement_window}
\end{equation}
This condition captures the basic information--backaction trade-off. Stronger monitoring provides more information per unit time, but reduces the coherent evolution time available before measurement-induced dephasing becomes dominant.

An important consequence of monitoring local \(Z_j\) observables is that Hamiltonian terms commuting with the monitored basis cannot be resolved from these signals at first order in time~\cite{Burgarth2014Identifiability}. In particular, terms whose action on every monitored site is proportional to \(\bbone\) or \(\sigma_z\) do not generate an instantaneous change in the measured coordinates. This limitation can be seen explicitly in Sec.~\ref{subsec:records_to_inverse_problems}. These coefficients must either be known in advance, accessed through additional measurement bases, or treated with higher-order reconstruction schemes. Throughout the main protocol, we focus on the Hamiltonian coefficients that are identifiable from the chosen weak-measurement signals.

Equation~\eqref{eq:wm_master_equation} is the dynamical model used throughout the protocol. The dephasing rates \(\Gamma_j\) are fixed by the measurement strength and are assumed to be known or independently calibrated~\cite{Schuster2005ACStark,PRXQuantum.4.040324}, so they are not additional learning targets. Our goal is to reconstruct the time-dependent Hamiltonian \(H(t)\) while consistently accounting for the measurement backaction induced by the monitoring process.

\subsection{Interaction graphs of Hamiltonians}
\label{subsec:interaction_graphs}

The locality structure of the Hamiltonian will be encoded through an interaction graph. This graph specifies which Hamiltonian terms overlap on common qubits and will be the central object used later to organize the reconstruction into local problems.

\begin{defn}[{Interaction graph and sparsely interacting Hamiltonian}]
Consider a Hamiltonian with known Pauli structure $\{W_i\}_{i \in [r]}$. Its interaction graph is the graph $\mathcal{G}=(V,E)$ whose vertices are the Hamiltonian terms,
\begin{equation}
    V=\{W_i\,:\,i=1,\ldots,r\},
\end{equation}
and whose edges connect pairs of terms with overlapping support,
\begin{equation}
    E=
    \left\{
    (W_i,W_j)\,:\,
    \operatorname{supp}(W_i)\cap\operatorname{supp}(W_j)\neq\varnothing,\,
    i\neq j
    \right\}.
\end{equation}
Here, \(\operatorname{supp}(W_i)\) denotes the set of qubits on which \(W_i\) acts nontrivially. We denote by
\begin{equation}
    \mathcal{D}=\max_{W_i\in V}\deg(W_i)
\end{equation}
the maximum degree of the interaction graph. A Hamiltonian is called \textit{sparsely interacting} if $\mathcal{D}=\mathcal{O}(1)$, that is, if each Hamiltonian term overlaps with only a bounded number of other terms independently of the total number of qubits.
\end{defn}

\begin{figure*}[t]
\centering
    \includegraphics[width=0.85\linewidth]{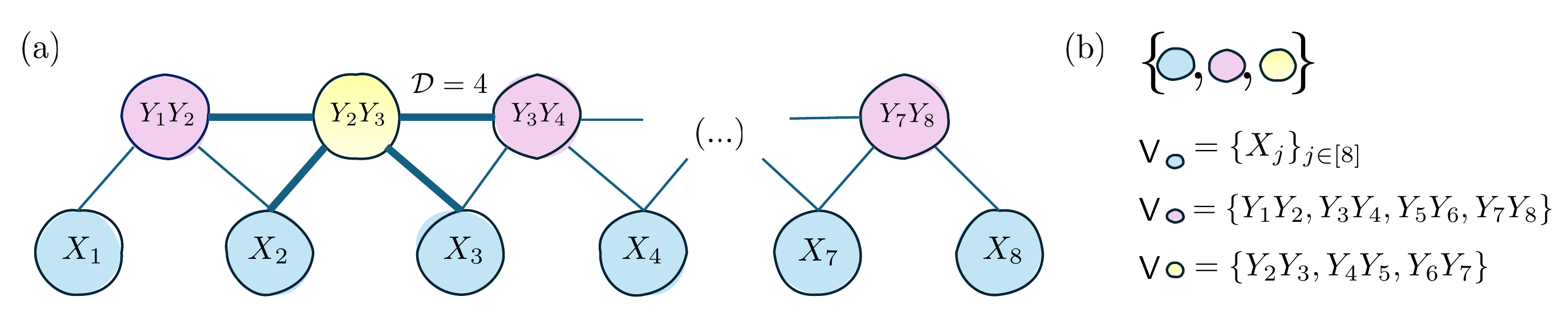}
    \caption{\justifying
    Interaction graph for the transverse field Ising model-like chain. 
    Vertices represent the Hamiltonian terms \(X_j\) and \(Y_jY_{j+1}\), and edges connect terms with overlapping support. 
    Panel (a) shows the graph structure, with maximum degree \(\mathcal D=4\). 
    Panel (b) gives a valid three-coloring, which partitions the terms into disjoint sets that can be addressed in parallel in the probe-construction step.
    }
    \label{fig:tfim_graph}
\end{figure*}

\begin{defn}[{Graph coloring}]
\label{def:graph_coloring}
Let $\mathcal{G}=(V,E)$ be a graph. A $C$-coloring of $\mathcal{G}$ is a map $c:V\rightarrow \{1,\ldots,C\}$ such that adjacent vertices have different colors, i.e. $c(v)\neq c(w)$ whenever $(v,w)\in E$. Equivalently, a coloring partitions the vertex set into independent sets
\begin{equation}
    V=\mathsf{V}_1\cup\cdots\cup \mathsf{V}_C,
\end{equation}
where no two vertices in the same subset $\mathsf{V}_c$ are connected by an edge. The minimum number of colors required to color $\mathcal{G}$ is the chromatic number, denoted by $\chi(\mathcal{G})$.
\end{defn}

The interaction graph provides the notion of locality used throughout the protocol. It does not require a specific spatial geometry; it only requires bounded overlap between Hamiltonian terms. In particular, geometrically local Hamiltonians on bounded-degree lattices are sparsely interacting in this sense, since increasing the number of qubits does not increase the number of terms overlapping with any fixed local interaction. Graph coloring then provides a way to organize this local structure into compatible sets: vertices with the same color correspond to Pauli terms with disjoint support and can therefore be addressed simultaneously by suitable probe configurations. Standard graph-coloring bounds imply that~\cite{DiestelGraphTheory}
\begin{equation}
    \chi(\mathcal{G})\leq \mathcal{D}+1.
\end{equation}
This observation will be used later to construct probe configurations whose number is controlled by the local connectivity of the Hamiltonian rather than by the total system size.

\begin{example}[Transverse field Ising-type chain]
\label{ex:tfim_graph}
In Fig.~\ref{fig:tfim_graph}, we show a sample interaction graph for an 8-qubit transverse field Ising model (TFIM), whose Hamiltonian is
\begin{equation}
    H_{\mathrm{TFIM}}
    =
    \sum_{j=1}^{7} J_j Y_jY_{j+1}
    +
    \sum_{j=1}^{8} h_j X_j .
\end{equation}
The TFIM will serve as a prototypical example for the rest of this work.

\end{example}

\section{Protocol description}
\label{sec:protocol}

\subsection{Overview}
Building on the weak-measurement model introduced in Sec.~\ref{subsec:weak_measurements}, we now describe the Hamiltonian reconstruction protocol. The input consists of the known Pauli structure $\{W_i\}_{i\in[r]}$, a set of probe preparations, calibrated dephasing rates $\{\Gamma_j\}_{j\in[n]}$, and empirical weak-measurement signals $\hat z_j(t_k)$ sampled at times $t_k=k\Delta t$. The output is an estimate $\hat{\boldsymbol{\mu}}(t_k)$ of the Hamiltonian coefficients at each reconstruction time.

The protocol alternates between two operations. First, numerical time derivatives of the measured signals are converted into linear inverse problems for the instantaneous Hamiltonian coefficients. Second, the reconstructed Hamiltonian is used together with the calibrated dephasing model to propagate the state estimates to the next time step. This dynamical update is required because the inversion matrices depend on expectation values of Pauli strings that are not directly measured.

The interaction graph introduced in Sec.~\ref{subsec:interaction_graphs} determines the locality of these inverse problems. Each local signal $\hat z_j(t_k)$ is sensitive only to Hamiltonian terms that do not commute with $Z_j$. We call this set the active neighborhood of site $j$. The protocol therefore decomposes the global reconstruction into local problems supported on these neighborhoods. Since the size of each active neighborhood is bounded by the local interaction degree, the number of probe configurations required for invertibility is controlled by $\mathcal{D}$ rather than by the total number of qubits. We now derive these equations explicitly.

\subsection{From weak-measurement signals to local inverse problems}
\label{subsec:records_to_inverse_problems}

The first step of the protocol is to convert the weak-measurement signals into a linear equation for the instantaneous Hamiltonian coefficients. This follows from the equation of motion of the monitored observables. Starting from the master equation~\eqref{eq:wm_master_equation}, the adjoint evolution of an arbitrary observable $A$ is
\begin{equation}
    \frac{d}{dt}\langle A\rangle_t
    =
    -i\langle [A,H(t)]\rangle_t
    +
    \sum_{j'=1}^{n}
    \frac{\Gamma_{j'}}{2}
    \left\langle
     D^{\dagger}[Z_{j'}]A
    \right\rangle_t ,
    \label{eq:adjoint_lindblad}
\end{equation}
where $\langle A\rangle_t=\mathrm{Tr}[A \rho(t)]$. For the monitored coordinate $A=Z_j$, the dissipative contribution vanishes, since $Z_j$ commutes with every monitored operator $Z_{j'}$. Hence
\begin{equation}
    \dot z_j(t)
    =
    -i\,\mathrm{Tr}\!\left([Z_j,H(t)]\rho(t)\right).
    \label{eq:zj_dynamics}
\end{equation}
The measurement backaction therefore does not appear explicitly in the differential equation for the measured coordinate. It remains relevant through the evolution of the state $\rho(t)$, which is propagated with the full master equation.

The signals are sampled at times $t_k=k\Delta t$. To estimate the time derivative in Eq.~\eqref{eq:zj_dynamics} from the sampled signals, we use a linear estimator with fixed weights $w_i$ on a finite set of neighboring time steps $I\subset\mathbb Z$,
\begin{equation}
\begin{gathered}
    d\Sigma_j(t_k)
    \coloneq
    \frac{1}{\Delta t}\sum_{i\in I} w_i\, z_j(t_{k+i}),
    \\
    \sum_{i\in I}w_i=0,
    \qquad
    \sum_{i\in I}i\,w_i=1.
\end{gathered}
    \label{eq:finite_difference_record}
\end{equation}
The two conditions make the estimator exact for signals that are linear in time. For general signals, we assume that it approximates the derivative up to a deterministic error of order $q\geq1$,
\begin{equation}
    \left|d\Sigma_j(t_k)-\dot z_j(t_k)\right|\leq B\,\Delta t^{q},
    \label{eq:derivative_bias}
\end{equation}
where the constant $B$ depends on the weights and on the higher time derivatives of the signals. Then Eq.~\eqref{eq:zj_dynamics} gives
\begin{equation}
    d\Sigma_j(t_k)
    =
    -i\,\mathrm{Tr}\!\left([Z_j,H(t_k)]\rho(t_k)\right)
    +
    \mathcal{O}(\Delta t^{q}).
    \label{eq:zj_finite_difference_ideal}
\end{equation}
The simplest choice is the forward finite difference, $w_0=-1$ and $w_1=1$, which gives $d\Sigma_j(t_k)=[z_j(t_{k+1})-z_j(t_k)]/\Delta t$ and $q=1$. Higher-order finite differences give $q>1$, and Savitzky--Golay derivative filters~\cite{SavitzkyGolay1964} are also of the form~\eqref{eq:finite_difference_record}. For any choice of weights, we refer to $d\Sigma_j(t_k)$ as the ideal finite-difference signal. In the experiment, $z_j(t_k)$ is replaced by the empirical average $\hat z_j(t_k)$, giving the empirical finite difference $d\hat\Sigma_j(t_k)$. The time step $\Delta t$ is a protocol parameter: decreasing it reduces the bias $B\Delta t^{q}$ but amplifies the statistical noise in the finite difference, a trade-off quantified in Sec.~\ref{sec:performance}.

Substituting the Hamiltonian ansatz~\eqref{eq:hamiltonian_ansatz} into Eq.~\eqref{eq:zj_finite_difference_ideal} yields a linear relation between the measured finite difference and the Hamiltonian coefficients,
\begin{equation}
    d\Sigma_j(t_k)
    =
    \sum_{m=1}^{r}
    M_{jm}(t_k)\,\mu_m(t_k)
    +
    \mathcal{O}(\Delta t^{q}),
    \label{eq:single_probe_linear_relation}
\end{equation}
where
\begin{equation}
    M_{jm}(t_k)
    =
    -i\,\mathrm{Tr}\!\left([Z_j,W_m]\rho(t_k)\right).
    \label{eq:single_probe_matrix_entries}
\end{equation}
For a single probe state, this gives only one equation per monitored site at each time step. Since the number of unknown coefficients is generally larger than the number of measured local signals, one must repeat the same experiment with different initial probe states. For instance, the transverse-field Ising chain of Example~\ref{ex:tfim_graph} already contains $2n-1$ coefficients.

Let $\mathcal{S}=\{\rho_s(0)\}_{s\in[S]}$ be a set of $S$ known probe states. Each probe evolves under the same Hamiltonian and produces a signal $ z_j^{(s)}(t_k)$. For each probe, we define
\begin{equation}
\begin{aligned}
    d\Sigma_j^{(s)}(t_k)
    &\coloneq
    \frac{1}{\Delta t}\sum_{i\in I} w_i\, z_j^{(s)}(t_{k+i}),
    \\
    M_{jm}^{(s)}(t_k)
    &=
    -i\,\mathrm{Tr}\!\left([Z_j,W_m]\rho_s(t_k)\right).
\end{aligned}
    \label{eq:finite_difference_probe_s}
\end{equation}
Stacking the $S$ probe states gives the central inversion equation of the protocol,
\begin{equation}
    d\boldsymbol{\Sigma}_j(t_k)
    =
    \matM_j(t_k)\boldsymbol{\mu}(t_k)
    +
    \mathcal{O}(\Delta t^{q})
    \label{eq:multiprobe_linear_relation}
\end{equation}
with
\begin{equation}
\begin{aligned}
    d\boldsymbol{\Sigma}_j(t_k)
    &=
    \left(
    d\Sigma_j^{(1)}(t_k),\ldots,d\Sigma_j^{(S)}(t_k)
    \right)^T,
    \\
    \left(\matM_j(t_k)\right)_{sm}
    &=
    M_{jm}^{(s)}(t_k).
\end{aligned}
    \label{eq:multiprobe_vector_matrix}
\end{equation}
Equation~\eqref{eq:multiprobe_linear_relation} is the object that turns weak-measurement data into Hamiltonian estimates. The vector on the left is obtained from the measured signals, while the matrix $\matM_j(t_k)$ is computed from the current state estimates. The probe states provide independent rows of this matrix.

At first sight, Eq.~\eqref{eq:multiprobe_linear_relation} appears to involve the full coefficient vector $\boldsymbol{\mu}(t_k)$. However, most columns of $\matM_j(t_k)$ vanish identically. A term $W_m$ contributes to the signal of site $j$ only if it fails to commute with $Z_j$. To make the locality explicit, we introduce the \textit{active neighborhood} of the monitored site $j$ as the set of Hamiltonian terms that contribute to the first-order evolution of $z_j(t)$,
\begin{equation}
    \mathcal N(j)
    \coloneq
    \left\{
    W_m \,:\, [Z_j,W_m]\neq 0
    \right\}.
    \label{eq:active_neighborhood}
\end{equation}
Equivalently, $\mathcal N(j)$ contains precisely those Pauli strings whose support includes site $j$ and whose local action on that site is $\sigma_x$ or $\sigma_y$. Terms outside $\mathcal N(j)$ commute with $Z_j$ and do not appear in the local weak-measurement equation.

The inverse problem for site $j$ can therefore be restricted to the coefficient vector supported on the active neighborhood,
\begin{equation}
    \boldsymbol{\mu}_{\mathcal N(j)}(t_k)
    =
    \left(\mu_m(t_k)\,:\, W_m\in\mathcal N(j)\right).
\end{equation}
Let $\matM_j^{\mathrm{loc}}(t_k)$ be the submatrix of $\matM_j(t_k)$ obtained by keeping only the columns indexed by $\mathcal{N}(j)$. This matrix has dimension $S\times |\mathcal{N}(j)|$, where each row corresponds to one probe state and each column to one locally visible Hamiltonian coefficient. The local reconstruction equation is then
\begin{equation}
    d\boldsymbol{\Sigma}_j(t_k)
    =
    \matM_j^{\mathrm{loc}}(t_k)
    \boldsymbol{\mu}_{\mathcal{N}(j)}(t_k)
    +
    \mathcal{O}(\Delta t^{q}).
    \label{eq:local_inverse_problem}
\end{equation}

To reconstruct the full Hamiltonian, the protocol solves Eq.~\eqref{eq:local_inverse_problem} for every monitored site and then combines the locally reconstructed coefficients. This is the central reduction of the protocol: instead of inverting a single global system involving all $r$ Hamiltonian parameters, we solve a collection of local inverse problems whose sizes are determined by the active neighborhoods $\mathcal{N}(j)$. Consequently, identifiability is a local condition. For each $j$, the matrix $\matM_j^{\mathrm{loc}}(t_k)$ must have full column rank, which requires
\begin{equation}
    S\geq |\mathcal{N}(j)|, 
    \qquad \forall j\in[n].
\end{equation}
When this condition is satisfied and the local matrix is well conditioned, the coefficients in the active neighborhood are estimated using the Moore--Penrose left pseudoinverse~\cite{BenIsraelGreville2003Pseudoinverse},
\begin{equation}
    \boldsymbol{\mu}_{\mathcal{N}(j)}(t_k)
    =
    \left(\matM_j^{\mathrm{loc}}(t_k)\right)^+
    d\boldsymbol{\Sigma}_j(t_k)
    \label{eq:local_pseudoinverse}
\end{equation}
up to the discretization error. Equation~\eqref{eq:local_pseudoinverse} is the practical estimator used by the protocol. The measured finite differences form the data vector, while the propagated state estimates determine the local reconstruction matrix.

{In the actual reconstruction, Eq.~\eqref{eq:local_pseudoinverse} is implemented with empirical finite differences and reconstructed state estimates, namely by replacing $d\boldsymbol{\Sigma}_j(t_k)$ and $\matM_j^{\mathrm{loc}}(t_k)$ with $d\hat{\boldsymbol{\Sigma}}_j(t_k)$ and $\hat{\matM}_j^{\mathrm{loc}}(t_k)$.}

The important consequence is that the required number of probe states is not set by the total number of Hamiltonian coefficients, but by the largest local inverse problem,
\begin{equation}
    S\geq \max_j |\mathcal{N}(j)|.
    \label{eq:probe_lower_bound}
\end{equation}
For sparsely interacting Hamiltonians, this quantity is controlled by the maximum degree $\mathcal{D}$ of the interaction graph. Indeed, all terms in $\mathcal{N}(j)$ act nontrivially on qubit $j$, so they have overlapping support and form a clique in the interaction graph. Hence $|\mathcal{N}(j)|-1\leq \mathcal{D}$, and therefore
\begin{equation}
    \max_j |\mathcal{N}(j)|\leq \mathcal{D}+1.
    \label{eq:active_neighborhood_degree_bound}
\end{equation}
Thus each local inverse problem requires at most $\mathcal D+1$ probe configurations. In Sec.~\ref{sec:probe_configurations}, we show that all local problems can be addressed simultaneously with $\mathcal O(\mathcal D^3)$ global probe configurations when the Pauli strings have bounded weight. This is the main locality advantage of the protocol. For sparsely interacting Hamiltonians, $\mathcal{D}$ remains bounded regardless of system size, so the number of probe configurations required to set up the local inversions is governed by the local connectivity of the Hamiltonian rather than by the total number of qubits or the total number of Hamiltonian terms.

\begin{figure*}[t]
    \includegraphics[width=0.88\textwidth]{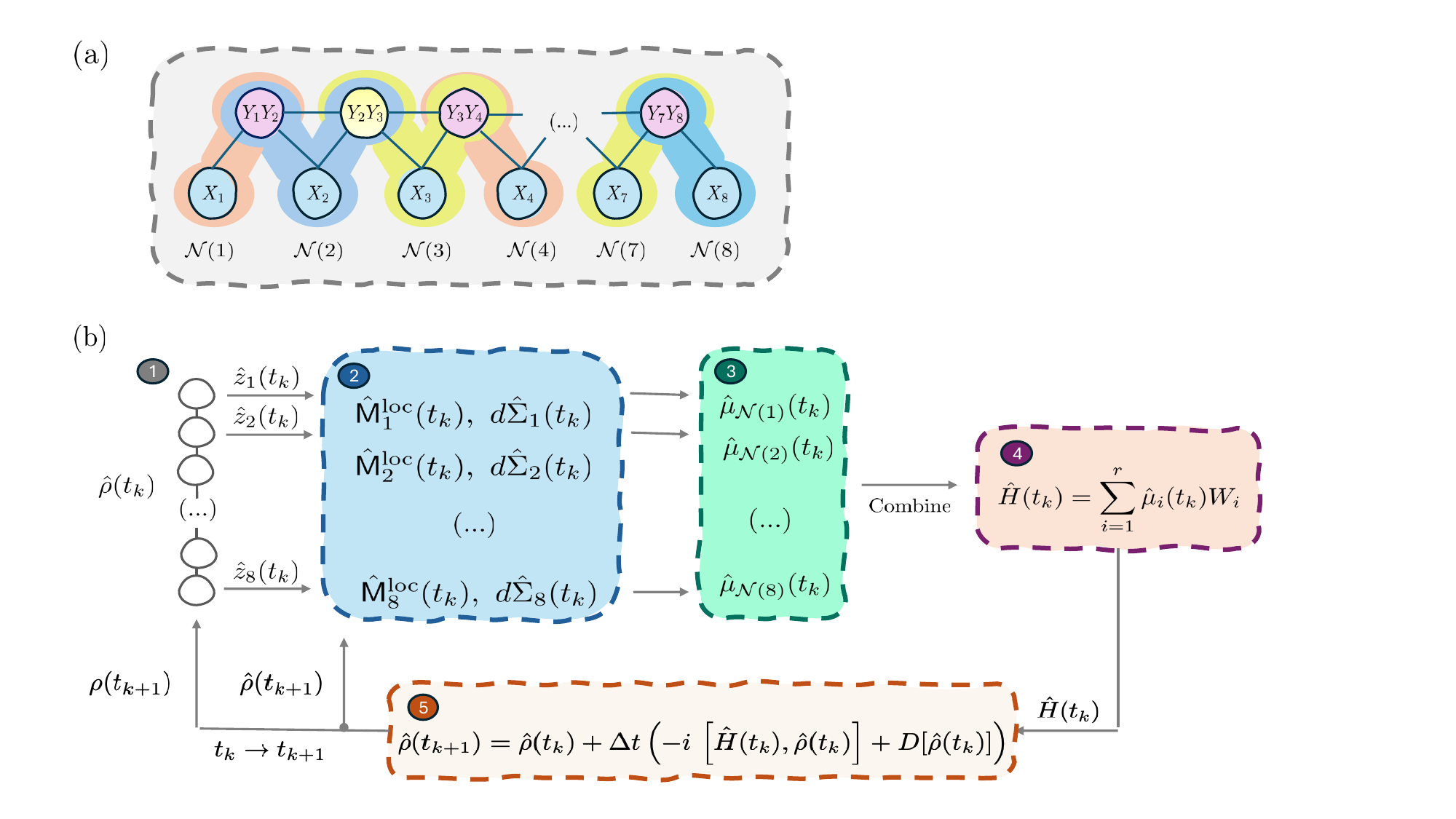}
    \caption{\justifying
    Local structure of the weak-measurement reconstruction protocol. 
    Panel (a) shows the active neighborhoods \(\mathcal N(j)\) for the TFIM-like chain, highlighted on the interaction graph. 
    Each active neighborhood contains the Hamiltonian terms contributing to the first-order evolution of the monitored signal at site \(j\). 
    Panel (b) sketches the reconstruction at a fixed time \(t_k\). 
    The weak-measurement signals \(\hat z_j(t_k)\) are converted into finite-difference data vectors \(d\hat{\boldsymbol{\Sigma}}_j(t_k)\), which define local inverse problems with matrices \(\hat{\matM}_j^{\mathrm{loc}}(t_k)\). 
    Solving these local problems gives estimates of the neighborhood coefficients \(\hat{\boldsymbol{\mu}}_{\mathcal N(j)}(t_k)\), which are then combined into the global Hamiltonian estimate \(\hat H(t_k)=\sum_i \hat\mu_i(t_k)W_i\).
    Finally, \(\hat H(t_k)\) propagates the probe-state estimates to \(\hat\rho(t_{k+1})\) through Eq.~\eqref{eq:state_update}, which provides the reconstruction matrices for the next time step.}
    \label{fig:reconstruction_protocol}
\end{figure*}

Figure~\ref{fig:reconstruction_protocol} illustrates the locality structure of the reconstruction for the transverse-field Ising example introduced in Sec.~\ref{subsec:interaction_graphs}. The figure is meant to show how the physical chain, the interaction graph, the active neighborhoods, and the local reconstructions fit together at a fixed time step.

\subsection{Hamiltonian reconstruction step-by-step}
\label{subsec:step_by_step_reconstruction}

We now summarize the reconstruction procedure. The required inputs are the known Pauli structure $\{W_m\}_{m\in[r]}$, the empirical weak-measurement signals $\hat z_j^{(s)}(t_k)$, the calibrated dephasing rates $\{\Gamma_j\}_{j\in[n]}$, and a set of known probe states $\mathcal{S}=\{\rho_s(0)\}_{s\in[S]}$. The active neighborhoods $\mathcal{N}(j)$ are determined directly from the Pauli structure through Eq.~\eqref{eq:active_neighborhood}. In the implementation considered here, the probe states can be chosen as pure separable states.

The algorithm proceeds as follows.

\begin{enumerate}
    \item \textit{Initialization.}  
    For each probe state, initialize the state estimate as
    \begin{equation}
        \hat\rho_s(0)=\rho_s(0),
        \qquad s\in[S].
    \end{equation}
    The initial states are assumed to be known, so the reconstruction starts from a fully specified estimate of each probe trajectory.

    \item \textit{Construct the local reconstruction matrices.}  
    At time $t_k$, use the current state estimates $\{\hat\rho_s(t_k)\}_{s\in[S]}$ to construct, for every monitored site $j$, the local matrix $\hat{\matM}_j^{\mathrm{loc}}(t_k)$ restricted to the active neighborhood $\mathcal{N}(j)$.

    \item \textit{Estimate the local coefficients.}  
    From the weak-measurement signals, form the data vector $d\hat{\boldsymbol{\Sigma}}_j(t_k)$ and solve the local inverse problem
    \begin{equation}
        \hat{\boldsymbol{\mu}}_{\mathcal{N}(j)}(t_k)
        =
        \left(\hat{\matM}_j^{\mathrm{loc}}(t_k)\right)^+
        d\hat{\boldsymbol{\Sigma}}_j(t_k).
        \label{eq:step_local_estimator}
    \end{equation}
    This inversion is performed independently for all monitored sites $j$.

    \item \textit{Assemble the Hamiltonian estimate.}  
    The local estimates $\hat{\boldsymbol{\mu}}_{\mathcal{N}(j)}(t_k)$ are combined into a single vector $\hat{\boldsymbol{\mu}}(t_k)$. If the same coefficient appears in more than one active neighborhood, the redundant estimates can be averaged or combined through a least-squares rule. The reconstructed coefficient vector defines the Hamiltonian estimate
    \begin{equation}
        \hat H(t_k)
        =
        \sum_{m=1}^{r}
        \hat\mu_m(t_k)W_m.
        \label{eq:estimated_hamiltonian_tk}
    \end{equation}

    \item \textit{Update the state estimates.}  
    Propagate each probe-state estimate to the next time step using the calibrated weak-measurement dynamics and an explicit integration scheme. For the Euler scheme,
    \begin{equation}
    \begin{aligned}
        \hat\rho_s(t_{k+1})
        &=
        \hat\rho_s(t_k)
        \\
        &+
        \Delta t
        \Big[
        -i[\hat H(t_k),\hat\rho_s(t_k)]\\
        &\qquad +
        \sum_{j=1}^{n}
        \frac{\Gamma_j}{2}
         D[Z_j]\hat\rho_s(t_k)
        \Big],
    \end{aligned}
        \label{eq:state_update}
    \end{equation}
    for all $s\in[S]$. More generally, any causal scheme can be used, that is, any update that uses $\hat H$ only at the reconstruction times $t_{k'}\leq t_k$ and the probe states at previous time steps (Definition~\ref{def:integrator} in Appendix~\ref{app:matrix_error_bound}). Examples are the exponential of the generator at $t_k$, Runge--Kutta methods with $\hat H$ extrapolated from previous reconstruction times, and Adams--Bashforth methods.
\end{enumerate}

Steps 2--6 are repeated for every time step in the reconstruction window. The protocol therefore alternates between local Hamiltonian estimation and state propagation. The measurement signals supply the data vectors, while the propagated state estimates supply the matrices needed for the next inversion. The output is the sequence of coefficient estimates $\{\hat{\boldsymbol{\mu}}(t_k)\}_k$, or equivalently the reconstructed Hamiltonian trajectory $\{\hat H(t_k)\}_k$.

\section{Graph-coloring separable-probe construction}
\label{sec:probe_configurations}

In Sec.~\ref{subsec:records_to_inverse_problems} we showed that the reconstruction reduces to local inverse problems supported on the active neighborhoods $\mathcal N(j)$. The rank condition for local invertibility requires enough probe configurations to make each local matrix $\matM_j^{\mathrm{loc}}(t_k)$ full column rank. In this section, we show two things. First, for Pauli Hamiltonians, this local rank condition can already be achieved using separable probe states. Second, these local probes can be systematically embedded into global separable configurations via a graph-coloring construction.

Initially, separable probes are sufficient because the local matrix entries are the expectations of Pauli strings. Indeed, for $W_m\in\mathcal N(j)$,
\begin{equation}
    \left(\matM_j^{\mathrm{loc}}(t_0)\right)_{s m}
    =
    -i\,\mathrm{Tr}\!\left([Z_j,W_m]\rho_s(t_0)\right).
\end{equation}
Since $W_m$ is a Pauli string and $[Z_j,W_m]\neq0$ for active terms, the operator
\begin{equation}
    Q_{jm}\coloneq -i[Z_j,W_m]
\end{equation}
is, up to an irrelevant constant, another Pauli string. Therefore, choosing a probe state with nonzero expectation value $\mathrm{Tr}(Q_{jm}\rho_s(t_0))$ activates the column associated with $W_m$. Such a response can be obtained with product states, since every Pauli string has separable eigenstates.

Activation of individual columns is not enough, however. To solve the local inverse problem, the probe set must produce linearly independent responses to all commutator strings $\{Q_{jm}:W_m\in\mathcal N(j)\}$. Thus, the local product probes are chosen so that the rows
\begin{equation}
    \left\{
    \mathrm{Tr}[Q_{jm}\rho_s(t_0)]
    \right\}_{W_m\in\mathcal N(j)}
\end{equation}
span the locally visible coefficient space. Such probes always exist. The strings $Q_{jm}$ with $W_m\in\mathcal N(j)$ are distinct Pauli strings supported on $\operatorname{supp}\mathcal N(j)$, hence linearly independent, and product states span the space of operators on this support. Therefore, the vectors $(\mathrm{Tr}[Q_{jm}\rho])_{W_m\in\mathcal N(j)}$ obtained from product states $\rho$ span $\mathbb R^{|\mathcal N(j)|}$, and $|\mathcal N(j)|$ of them can be chosen linearly independent. Since the local matrix for site $j$ has $|\mathcal N(j)|$ columns, the construction uses $|\mathcal N(j)|$ independent local product probes for that neighborhood. To apply the same preparation scheme uniformly across all sites, we take the worst-case number $\max_j|\mathcal N(j)|$. With this choice, $\matM_j^{\mathrm{loc}}(t_0)$ can be made full column rank for every monitored site, so entangled probes are not required to initialize the local inversions.

This statement is made at the initial time because the later reconstruction matrices depend on the propagated state estimates, which are generated iteratively by the protocol. The role of the probe design is therefore to guarantee that the local inversions are well initialized. Once the reconstruction starts, the matrices at later times are obtained from the dynamical update described in Sec.~\ref{subsec:step_by_step_reconstruction}.

The remaining question is how to impose these local separable probes on the full many-body system without preparing each active neighborhood independently. The guiding principle is parallelization. Local probes associated with different active neighborhoods can be implemented in the same global preparation whenever the corresponding neighborhoods act on disjoint sets of physical qubits. Since $\mathcal N(j)$ denotes a set of Hamiltonian terms, we associate with it the physical support
\begin{equation}
    \operatorname{supp}\mathcal N(j)
    =
    \bigcup_{W_m\in\mathcal N(j)}
    \operatorname{supp}(W_m).
    \label{eq:active_neighborhood_support}
\end{equation}
This is the set of qubits on which the local probe for the inverse problem at site $j$ must be specified.

\begin{defn}[Active-neighborhood graph]
\label{def:active_neighborhood_graph}
The \textit{active-neighborhood graph} is the graph
$\mathcal G_{\mathcal N}=(V_{\mathcal N},E_{\mathcal N})$ whose vertices are the active neighborhoods,
\begin{equation}
    V_{\mathcal N}
    =
    \{\mathcal N(j):j\in[n]\},
\end{equation}
and whose edges connect active neighborhoods with overlapping physical support,
\begin{equation}
\begin{aligned}
    &\bigl(\mathcal N(j),\mathcal N(j')\bigr)\in E_{\mathcal N}
    \\
    &\quad\Longleftrightarrow\quad
    \operatorname{supp}\mathcal N(j)
    \cap
    \operatorname{supp}\mathcal N(j')
    \neq \varnothing,
    \quad j\neq j'.
\end{aligned}
\end{equation}
A proper coloring of $\mathcal G_{\mathcal N}$ partitions the active neighborhoods into color classes whose physical supports are pairwise disjoint.
\end{defn}

A color class of $\mathcal G_{\mathcal N}$ can therefore be implemented by a single global separable probe state. If two active neighborhoods have the same color, their supports are disjoint, and the corresponding local probe choices act on different tensor factors of the full Hilbert space. This is precisely the condition needed to prepare those local probes simultaneously.

\begin{proposition}[Parallel separable-probe construction]
\label{prop:parallel_probe_construction}
Let $\mathcal G_{\mathcal N}$ be the active-neighborhood graph and let $\chi(\mathcal G_{\mathcal N})$ be its chromatic number. Let $\alpha^\mathrm{loc}=\max_j |\cN(j)|$ be the worst-case number of local product-state probes used to make the local reconstruction matrices full column rank at $t_0$. Then there exists a set of global separable probe configurations $\mathcal S$ satisfying
\begin{equation}
    |\mathcal S|
    \leq
    \chi(\mathcal G_{\mathcal N})\,
    \alpha^\mathrm{loc},
    \label{eq:probe_bound_conflict_coloring}
\end{equation}
such that all local inverse problems are solvable at the initial time.
\end{proposition}

The construction follows directly from the coloring. Let
\begin{equation}
    V_{\mathcal N}
    =
    \mathsf{V}_1\cup\cdots\cup \mathsf{V}_{\chi(\mathcal G_{\mathcal N})}
\end{equation}
be a proper coloring of $\mathcal G_{\mathcal N}$. For each color class $\mathsf{V}_c$, the supports of the active neighborhoods in $\mathsf{V}_c$ are pairwise disjoint. Hence, for every local probe label $a\in[\alpha^\mathrm{loc}]$, one can define the global product state
\begin{equation}
    \rho_c^{(a)}
    =
    \left(
    \bigotimes_{\mathcal N(j)\in \mathsf{V}_c}
    \rho_{\mathcal N(j)}^{(a)}
    \right)
    \otimes
    \rho_{\mathrm{ref}}^{(c)} ,
    \label{eq:global_probe_state_color}
\end{equation}
where $\rho_{\mathcal N(j)}^{(a)}=\bigotimes_{j'\in\operatorname{supp}\cN(j)} \rho_{j'}^{(a)}$ is the $a$-th local probe state on $\operatorname{supp}\mathcal N(j)$, and $\rho_{\mathrm{ref}}^{(c)}$ is a fixed reference state on all qubits outside
\begin{equation}
    \operatorname{supp}(\mathsf{V}_c)
    =
    \bigcup_{\mathcal N(j)\in \mathsf{V}_c}
    \operatorname{supp}\mathcal N(j).
\end{equation}
The tensor product in Eq.~\eqref{eq:global_probe_state_color} is well defined because active neighborhoods in the same color class have disjoint physical support. Repeating this construction for every color $c$ gives at most $\chi(\mathcal G_{\mathcal N}) \,\alpha^\mathrm{loc}$ global probe configurations.

The two factors in Eq.~\eqref{eq:probe_bound_conflict_coloring} play different roles. The factor $\alpha^\mathrm{loc}$ is the number of probes needed to solve the largest local inverse problem, whereas $\chi(\mathcal G_{\mathcal N})$ is the number of global preparations needed to run all local problems in parallel. The local factor needs no additional coloring argument, since it is already controlled by the interaction graph. All terms in $\mathcal N(j)$ act on site $j$, so they are pairwise adjacent in $\mathcal G$ and must receive different colors in any coloring of $\mathcal G$. Therefore,
\begin{equation}
    \alpha^\mathrm{loc}
    \leq
    \chi(\mathcal G)
    \leq
    \mathcal D+1.
    \label{eq:alpha_loc_bound}
\end{equation}
For the TFIM chain of Ex.~\ref{ex:tfim_graph}, $\alpha^\mathrm{loc}=\chi(\mathcal G)=3$, while $\mathcal D+1=5$. The global factor is controlled by $\mathcal D$ once the weight of the Pauli strings is bounded. If each $W_i$ acts nontrivially on at most $K$ qubits, any active neighborhood overlapping $\operatorname{supp}\mathcal N(j)$ contains one of the at most $(\mathcal D+1)^2$ terms that coincide with or overlap a term of $\mathcal N(j)$, and each such term belongs to at most $K$ active neighborhoods. Hence $\chi(\mathcal G_{\mathcal N})\leq K(\mathcal D+1)^2$ and $|\mathcal S|\leq K(\mathcal D+1)^3$, independently of $n$. This worst-case estimate is loose for regular geometries: for the TFIM chain, $\chi(\mathcal G_{\mathcal N})=3$ and the construction uses $9$ probe configurations.

\begin{figure*}[t]
    \includegraphics[width=0.85\linewidth]{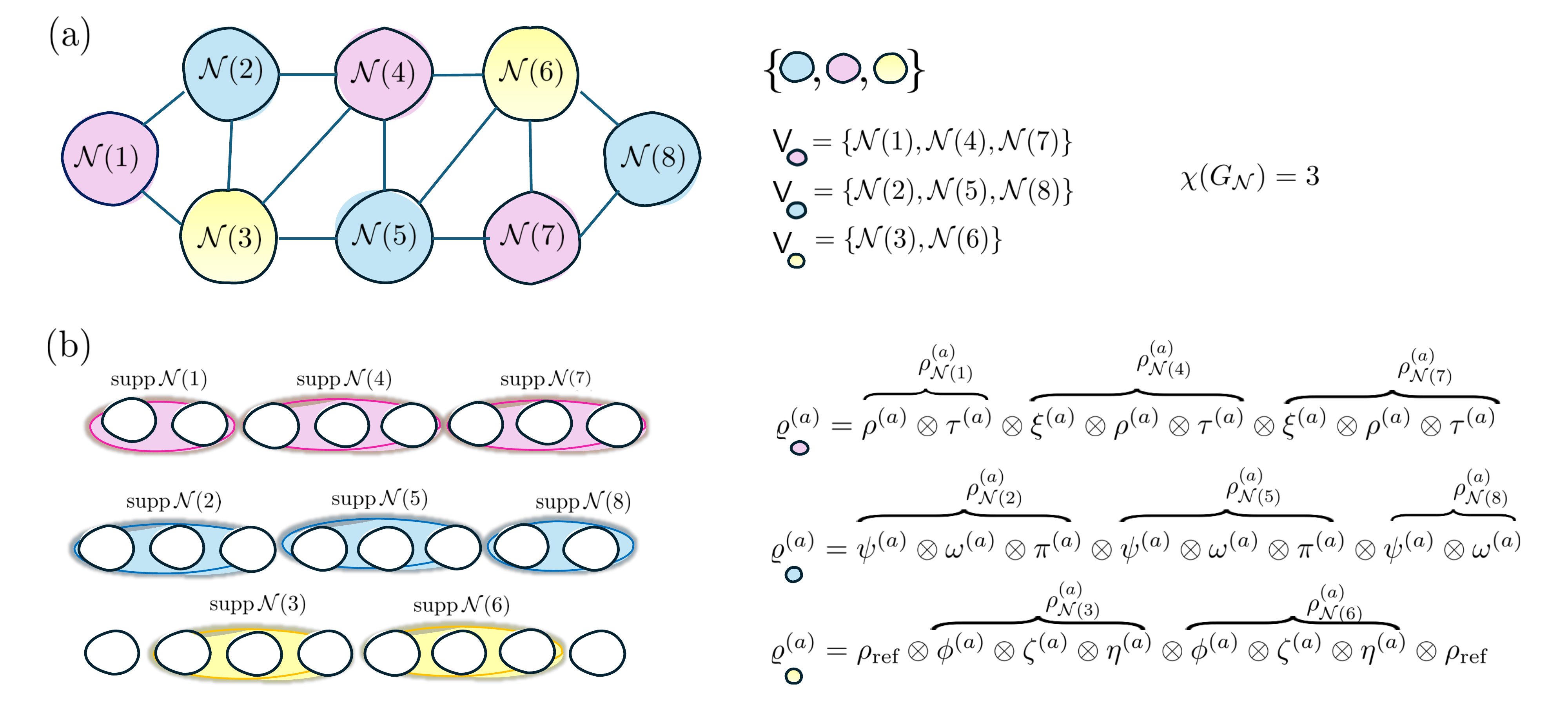}
    \caption{\justifying
    Graph-coloring construction of separable probe configurations for the TFIM-like chain. 
    Panel (a) shows the active-neighborhood graph \(\mathcal G_{\mathcal N}\), whose vertices are the local inverse problems \(\mathcal N(j)\). 
    The three-coloring \(\chi(\mathcal G_{\mathcal N})=3\) groups neighborhoods with disjoint physical support. 
    Panel (b) displays these supports on the physical qubit chain and shows the corresponding global product probes: neighborhoods with the same color are prepared simultaneously, while spectator qubits are assigned fixed reference states. 
    Repeating the construction for the local probe labels \(a\) gives the separable ensemble used to initialize all local inversions.
    }
    \label{fig:tfim_probe_coloring}
\end{figure*}

Whenever possible, the same local probe pattern can be reused across active neighborhoods belonging to the same color class. This happens when the Hamiltonian has a repeated local structure, as in translation-invariant or periodic chains. In that case, for all $\mathcal N(j),\mathcal N(j')\in \mathsf{V}_c$ with the same local geometry, one may choose
\begin{equation}
    \rho_{\mathcal N(j)}^{(a)}
    =
    \rho_{\mathcal N(j')}^{(a)}
\end{equation}
up to the natural relabeling of qubits. Thus, a single local probe design can be copied along the color class, reducing the practical overhead of constructing the probe ensemble. Figure~\ref{fig:tfim_probe_coloring} illustrates this for the TFIM chain, where the active-neighborhood graph admits a three-coloring and the probe configurations repeat periodically along the system.

The construction therefore provides a systematic separable-state preparation scheme adapted to the Hamiltonian structure. In its most economical form, one uses the minimal coloring of the active-neighborhood graph together with the smallest local product-probe set that makes each local matrix full rank. The resulting number of global probes is bounded by the coloring construction; it is not a lower bound over all possible probe strategies. Model-specific symmetries may reduce the number of useful preparations, while more redundant choices may improve stability.

This distinction matters experimentally. The minimal graph-coloring ensemble is designed to initialize all local inversions with a small separable overhead and is expected to perform well over short reconstruction windows. At longer times, however, this minimal ensemble can be less robust. Since it is tailored to the coloring, different qubits may play asymmetric roles across the preparations. As the probe states evolve coherently, some local reconstruction matrices can then become poorly conditioned, which amplifies statistical fluctuations and the errors accumulated in the state propagation. For this reason, in the numerical analysis we also consider an overcomplete separable ensemble obtained by preparing each qubit independently in an eigenstate of one of the Pauli operators,
\begin{equation}
    \rho
    =
    \bigotimes_{j=1}^{n}
    \ket{\alpha_j}\!\bra{\alpha_j},
    \qquad
    \ket{\alpha_j}\in\{\ket{x_\pm},\ket{y_\pm},\ket{z_\pm}\},
\end{equation}
with each local state drawn independently and uniformly. This overcomplete choice increases the number of probe configurations, but provides a more balanced sampling of the local operator space. In Sec.~\ref{sec:numerics}, we compare the performance of the minimal graph-coloring ensemble with this overcomplete Pauli-eigenstate ensemble numerically.

\section{Error bounds and sample complexity}
\label{sec:performance}

We now quantify the reconstruction accuracy of the weak-measurement Hamiltonian reconstruction protocol. The section has two goals. First, we bound the error in the reconstructed Hamiltonian coefficients at a fixed reconstruction time $t_k$. This bound separates the statistical error coming from the finite weak-measurement signals from the deterministic bias, which arises from the derivative approximation and the state propagation. Second, we use this estimate to derive a sample-complexity theorem for the number of trajectories required to reconstruct the Hamiltonian coefficients up to a prescribed accuracy. Detailed derivations of the bounds are given in Appendix~\ref{app:error_bounds}.

\begin{proposition}[Coefficient reconstruction error]
\label{prop:coefficient_error_bound}
Fix a reconstruction time $t_k$ and suppose that all local reconstruction matrices $\hat{\matM}_j^{\mathrm{loc}}(t_k)$ have full column rank. Suppose that the derivatives are estimated with Eq.~\eqref{eq:finite_difference_record}, with bias $B\Delta t^{q}$, and that the probe states are propagated with a causal scheme of order $p$ (Definition~\ref{def:integrator}). Let $S$ be the number of probe configurations and let $N_s$ be the number of trajectories used to estimate each averaged weak-measurement signal. Define $\kappa_w\coloneq\|w\|_1+\|w\|_2/\sqrt{2\Gamma_{\min}\Delta t}$, with $\Gamma_{\min}=\min_j\Gamma_j$, and
\begin{equation}
\begin{aligned}
    \sigma_{\min}(\hat{\matM})
    &=
    \min_{j\in[n]}
    \sigma_{\min}\!\left(\hat{\matM}_j^{\mathrm{loc}}(t_k)\right),
    \\
    \mu_{\max}
    &=
    \sup_{t\leq t_k}
    \max_{j\in[n]}
    \left\|
    \boldsymbol{\mu}_{\mathcal N(j)}(t)
    \right\|_2,
\end{aligned}
\end{equation}
where $\sigma_{\min}(A)$ denotes the smallest singular value of $A$, and
\begin{equation}
    \Delta H_{\max}(t_k)
    =
    \max_{0\leq k'<k}
    \left\|
    \hat H(t_{k'})-H(t_{k'})
    \right\|_\infty,
\end{equation}
with $\Delta H_{\max}(t_0)=0$, and $\bar L=L+2\Delta H_{\max}(t_k)$.
Then, with probability at least $1-\delta$, the reconstructed coefficient vector satisfies
\begin{equation}
\begin{aligned}
    \left\|
    \hat{\boldsymbol{\mu}}(t_k)
    -
    \boldsymbol{\mu}(t_k)
    \right\|_2
    &\leq
    \frac{\sqrt n}{\sigma_{\min}(\hat{\matM})}
    \Bigg[\frac{\kappa_w}{\Delta t}\sqrt{ \frac{2S}{N_s}\log\!\left(\frac{2Sn}{\delta}\right)
    }
    \\
    &+\sqrt{S}\,B\Delta t^{q}
    +
    2\sqrt{S(\mathcal D+1)}\,
    \frac{e^{\gamma\bar Lt_k}-1}{\bar L}
    \\
    &\quad\times
    \left(
    2\Delta H_{\max}(t_k)+C\Delta t^{p}
    \right)
    \mu_{\max}
    \Bigg].
\end{aligned}
\label{eq:coefficient_error_bound}
\end{equation}
Here $L$ is the trace-norm Lipschitz constant of the effective weak-measurement generator, $\bar L$ bounds the Lipschitz constant of the reconstructed generator used in the update, $C$ controls the truncation error of the integration scheme, and $\gamma\geq1$ quantifies how the scheme propagates state and Hamiltonian errors, as specified in Appendix~\ref{app:matrix_error_bound}. For the Euler update, $p=\gamma=1$.
\end{proposition}
Proposition~\ref{prop:coefficient_error_bound} is a per-time-step guarantee rather than an a priori one: $\sigma_{\min}(\hat{\matM})$ and $\Delta H_{\max}(t_k)$ depend on the reconstruction history up to $t_k$ and cannot be fixed before running the protocol, but, given this history, the bound certifies the accuracy of the coefficients reconstructed at $t_k$. The failure probability $\delta$ then refers only to the concentration of the empirical weak-measurement signals used to construct the finite-difference data vectors at $t_k$, and the remaining terms in Eq.~\eqref{eq:coefficient_error_bound} are deterministic. This distinction is important because the two contributions in Eq.~\eqref{eq:coefficient_error_bound} behave differently under changes of the numerical and statistical resources.

The first term is statistical and comes from estimating time derivatives through finite differences. At fixed $N_s$, it grows as $\Delta t$ decreases, since the finite difference amplifies the sampling noise in the measured signals. This noise enters through $\kappa_w$. The homodyne current has Gaussian noise of variance $1/(2\Gamma_j\Delta t)$, so weaker monitoring requires more trajectories. The remaining terms are deterministic. The term $\sqrt S\,B\Delta t^{q}$ is the bias of the derivative estimator, while the last term contains the error introduced by the state update. In particular, the integration scheme contributes $\mathcal O(\Delta t^{p})$, which reduces to $\mathcal O(\Delta t)$ for the Euler update, while $\Delta H_{\max}(t_k)$ accounts for the coefficient errors accumulated from previous reconstruction steps and fed back into the propagation of the probe states. Thus, $\Delta t$ cannot be chosen arbitrarily small or arbitrarily large; it must balance the statistical cost of estimating derivatives against the systematic errors introduced by the derivative estimator and the dynamical update.

This also clarifies the role of the number of trajectories. Increasing $N_s$ reduces only the statistical contribution associated with the empirical signals. It does not remove the deterministic bias coming from the derivative estimator, the time discretization, and the iterative state propagation. Consequently, for a fixed reconstruction window and a fixed time step, the coefficient error is expected to decrease with $N_s$ only until the sampling noise becomes comparable to the deterministic contribution. Beyond that point, the reconstruction reaches a bias floor that can only be lowered by improving the derivative estimator or the time integration, changing $\Delta t$, or shortening the reconstruction window, but not by increasing the number of trajectories.

We now give an outline of the proof; the detailed derivation is given in Appendix~\ref{app:error_bounds}. For a fixed monitored site $j$, the local estimator satisfies the perturbative stability bound
\begin{equation}
\begin{aligned}
&\left\|
\hat{\boldsymbol{\mu}}_{\mathcal N(j)}(t_k)
-
\boldsymbol{\mu}_{\mathcal N(j)}(t_k)
\right\|_2
\\
&\quad\leq
\left\|
\left(
\hat{\matM}_j^{\mathrm{loc}}(t_k)
\right)^+
\right\|_{\infty}
\Big(
\left\|
d\hat{\boldsymbol{\Sigma}}_j(t_k)
-
d\boldsymbol{\Sigma}_j(t_k)
\right\|_2
\\
&\qquad
+
\left\|
d\boldsymbol{\Sigma}_j(t_k)
-
\matM_j^{\mathrm{loc}}(t_k)
\boldsymbol{\mu}_{\mathcal N(j)}(t_k)
\right\|_2
\\
&\qquad
+
\left\|
\hat{\matM}_j^{\mathrm{loc}}(t_k)
-
\matM_j^{\mathrm{loc}}(t_k)
\right\|_{\infty}
\left\|
\boldsymbol{\mu}_{\mathcal N(j)}(t_k)
\right\|_2
\Big).
\end{aligned}
\label{eq:local_mu_error_perturbative}
\end{equation}
The first term is controlled by the concentration of the empirical signals. For a fixed monitored site $j$, the $S$ probe configurations have already been stacked into the vector $d\hat{\boldsymbol{\Sigma}}_j(t_k)$, so Hoeffding concentration gives
\begin{equation}
    \left\|
    d\hat{\boldsymbol{\Sigma}}_j(t_k)
    -
    d\boldsymbol{\Sigma}_j(t_k)
    \right\|_2
    \leq
    \frac{\kappa_w}{\Delta t}
    \sqrt{
    \frac{2S}{N_s}
    \log\!\left(
    \frac{2S}{\delta}
    \right)
    }.
    \label{eq:finite_difference_concentration_bound_local}
\end{equation}
A union bound over all monitored sites gives the logarithmic factor $\log(2Sn/\delta)$ appearing in Eq.~\eqref{eq:coefficient_error_bound}.

The second term is the bias of the derivative estimator. By Eqs.~\eqref{eq:zj_dynamics} and~\eqref{eq:derivative_bias}, each of its $S$ entries is bounded by $B\Delta t^{q}$, so it is at most $\sqrt S\,B\Delta t^{q}$.

The third term is controlled by the perturbation of the reconstruction matrices. Since the entries of $\matM_j^{\mathrm{loc}}$ are expectation values of Pauli commutators evaluated on the propagated states, the matrix error is bounded by the trace-norm distance between the exact and reconstructed state trajectories. This gives
\begin{equation}
\begin{aligned}
    \left\|
    \hat{\matM}_j^{\mathrm{loc}}(t_k)
    -
    \matM_j^{\mathrm{loc}}(t_k)
    \right\|_{\infty}
    &\left\|
    \boldsymbol{\mu}_{\mathcal N(j)}(t_k)
    \right\|_2
    \\
    \leq{}&
    2\sqrt{S(\mathcal D+1)}\,
    \frac{e^{\gamma\bar Lt_k}-1}{\bar L}
    \\
    &\times
    \left(
    2\Delta H_{\max}(t_k)+C\Delta t^{p}
    \right)
    \mu_{\max}.
\end{aligned}
    \label{eq:matrix_error_bound}
\end{equation}
Finally, using
\begin{equation}
    \left\|
    \left(
    \hat{\matM}_j^{\mathrm{loc}}(t_k)
    \right)^+
    \right\|_{\infty}
    =
    \frac{1}{
    \sigma_{\min}\!\left(
    \hat{\matM}_j^{\mathrm{loc}}(t_k)
    \right)
    },
\end{equation}
and combining the local estimates over all monitored sites gives the prefactor $\sqrt n$ and the global conditioning factor $\sigma_{\min}(\hat{\matM})^{-1}$ in Eq.~\eqref{eq:coefficient_error_bound}.

The following theorem rewrites Proposition~\ref{prop:coefficient_error_bound} as a sufficient trajectory requirement. It should be understood as an achievability statement: if the deterministic bias is below the target accuracy and if the number of trajectories satisfies the stated bound, then the desired reconstruction accuracy follows with high probability. The bound is not claimed to be necessary.

Define the deterministic bias term
\begin{equation}
\begin{aligned}
    B_{\mathrm{det}}(t_k,\Delta t)
    \coloneq{}&
    \sqrt{S}\,B\Delta t^{q}
    +
    2\sqrt{S(\mathcal D+1)}\,
    \frac{e^{\gamma\bar Lt_k}-1}{\bar L}
    \\
    &\times
    \left(
    2\Delta H_{\max}(t_k)+C\Delta t^{p}
    \right)
    \mu_{\max}.
\end{aligned}
    \label{eq:deterministic_bias}
\end{equation}

\begin{thm}[Sample complexity]
\label{thm:sample_complexity}
Fix an accuracy $\epsilon>0$ and a failure probability $\delta\in(0,1)$. Suppose that
\begin{equation}
    \epsilon
    >
    \frac{\sqrt n}{\sigma_{\min}(\hat{\matM})}
    B_{\mathrm{det}}(t_k,\Delta t).
    \label{eq:sample_complexity_bias_condition}
\end{equation}
Then it is sufficient to take
\begin{equation}
    N_s
    \geq
    \frac{2\kappa_w^2Sn}{\Delta t^2}
    \frac{
    \log\!\left(
    \frac{2Sn}{\delta}
    \right)
    }{
    \left(
    \epsilon\sigma_{\min}(\hat{\matM})
    -
    \sqrt n\,B_{\mathrm{det}}(t_k,\Delta t)
    \right)^2
    }
    \label{eq:sample_complexity_bound}
\end{equation}
trajectories per probe configuration to guarantee
\begin{equation}
    \left\|
    \hat{\boldsymbol{\mu}}(t_k)
    -
    \boldsymbol{\mu}(t_k)
    \right\|_2
    \leq
    \epsilon
\end{equation}
with probability at least $1-\delta$.
\end{thm}

The condition in Eq.~\eqref{eq:sample_complexity_bias_condition} states that the requested accuracy must be larger than the deterministic bias, which arises from the derivative approximation and the state propagation. This is necessary for the bound to be informative, since increasing $N_s$ only reduces the statistical contribution and cannot remove the bias caused by the derivative estimator, the state propagation, and the time discretization. Once this condition is satisfied, Eq.~\eqref{eq:sample_complexity_bound} gives a sufficient number of trajectories per probe configuration. Both contributions are worst-case estimates. In particular, the deterministic term grows as $e^{\gamma\bar Lt_k}$, where $L$ is the worst-case Lipschitz constant of the generator, of order $2\|H\|_\infty$. This term is therefore informative only for reconstruction times $t_k\lesssim1/(\gamma\bar L)$.

\vspace{4mm}
\begin{main_res}
Following the algorithm detailed in Sec.~\ref{subsec:step_by_step_reconstruction}, to achieve with high probability a global reconstruction precision on the Hamiltonian coefficients of order $\mathcal O(\epsilon)$,
it suffices, in the regime where the deterministic bias is negligible compared with the target accuracy, to use
\begin{equation}\notag
    \mathcal O\!\left(
    \frac{\kappa_w^2Sn}{\Delta t^2\epsilon^2\sigma_{\min}(\hat{\matM})^2}
    \log\!\left(
    \frac{2Sn}{\delta}
    \right)
    \right)
\end{equation}
trajectories per probe configuration. For sparsely interacting Hamiltonians, the number of probe configurations $S$ is controlled by the local interaction structure. Thus, in this regime, the trajectory requirement is governed by local connectivity, the conditioning of the reconstruction matrices, the time resolution $\Delta t$, and the desired global accuracy.
\end{main_res}

\section{Numerical results}
\label{sec:numerics}

We now numerically validate the reconstruction protocol for time-dependent spin Hamiltonians with up to $n=8$ qubits. The goal is threefold: first, to show that weak-measurement signals allow one to reconstruct Hamiltonian coefficients as functions of time; second, to compare the minimal graph-coloring probe ensemble with overcomplete separable probe ensembles; and third, to verify the sample-complexity scaling predicted by Theorem~\ref{thm:sample_complexity}. Since the learning task is performed at each reconstruction time step, we compare the reconstructed Hamiltonian directly with the target coefficient trajectories.

Throughout this section, we use the TFIM-like nearest-neighbor chain introduced in Ex.~\ref{ex:tfim_graph}, now with smoothly time-dependent coefficients,
\begin{equation}
    H(t)
    =
    \sum_{j=1}^{n}
    h_j(t)X_j
    +
    \sum_{j=1}^{n-1}
    J_j(t)Y_jY_{j+1}.
    \label{eq:numerical_tfim_hamiltonian}
\end{equation}
We set $\hbar=1$ and choose a reference frequency $\omega_0$ such that the dimensionless local-field coefficients satisfy $h_j(t)=O(1)$. The Hamiltonian coefficients and dephasing rates are then expressed in units of $\omega_0$, while time is measured in units of $\omega_0^{-1}$.
The reconstruction is performed over the time window $t_{\mathrm{rec}}=3.0$, chosen to remain within the transient regime set by the measurement-induced dephasing scale $1/\Gamma_d$. In this regime, the measurement signals retain coherent information about the Hamiltonian before dephasing dominates the dynamics. The signals are simulated with a time discretization of the weak-measurement model of Sec.~\ref{subsec:weak_measurements}, with $\Gamma_j=\Gamma_d$ for all qubits, as detailed in Appendix~\ref{app:numerical_details}. The value $\Gamma_d=10^{-3}$ is conservative. When the Gaussian part of $\kappa_w$ dominates, Theorem~\ref{thm:sample_complexity} gives $N_s\propto1/\Gamma_{\min}$, so the number of trajectories can be reduced by choosing the largest measurement strength compatible with Eq.~\eqref{eq:weak_measurement_window}.

\textit{Time-dependent coefficient reconstruction.}
We first show a representative reconstruction for a chain with $n=6$ qubits, dephasing rate $\Gamma_d=10^{-3}$, an overcomplete separable Pauli-eigenstate ensemble with $S=36$ probe configurations, and an effective trajectory number $N_s=10^9$ per probe configuration. The reconstruction time step is chosen based on the trade-off discussed in Sec.~\ref{sec:performance}. Decreasing $\Delta t$ improves the integration scheme but amplifies finite-difference statistical noise, while increasing $\Delta t$ reduces this noise at the cost of a larger integration bias. We therefore use a balanced time step with scaling
\begin{equation}\label{eq:time_step}
    \Delta t_{\mathrm{opt}}
    \propto
    \left[
    \frac{
    \kappa_w^2\log\!\left(
    \frac{2Sn}{\delta}
    \right)
    }{
    N_s(\mathcal D+1)
    }
    \right]^{\frac{1}{2(\bar p+1)}},
\end{equation}
with $\bar p=\min\{p,q\}$. This scaling balances the statistical term in Eq.~\eqref{eq:coefficient_error_bound}, of order $\kappa_w/\Delta t$, against the deterministic bias, of order $\Delta t^{\bar p}$. Since $\kappa_w$ depends on $\Delta t$, Eq.~\eqref{eq:time_step} is implicit. When the bounded part of $\kappa_w$ dominates, $\kappa_w$ is constant. When its Gaussian part dominates, $\kappa_w^2\propto1/(\Gamma_{\min}\Delta t)$ and the exponent becomes $1/(2\bar p+3)$. In our simulations, the probe states are propagated with the Euler update, so that $p=1$ and hence $\bar p=1$. This trade-off applies to estimators with fixed weights. For a filter whose window is fixed in time, as in our simulations, the statistical term does not depend on $\Delta t$ and the trade-off is controlled by the window width instead.

\begin{figure*}[t]
    \centering
    \includegraphics[width=0.8\textwidth]{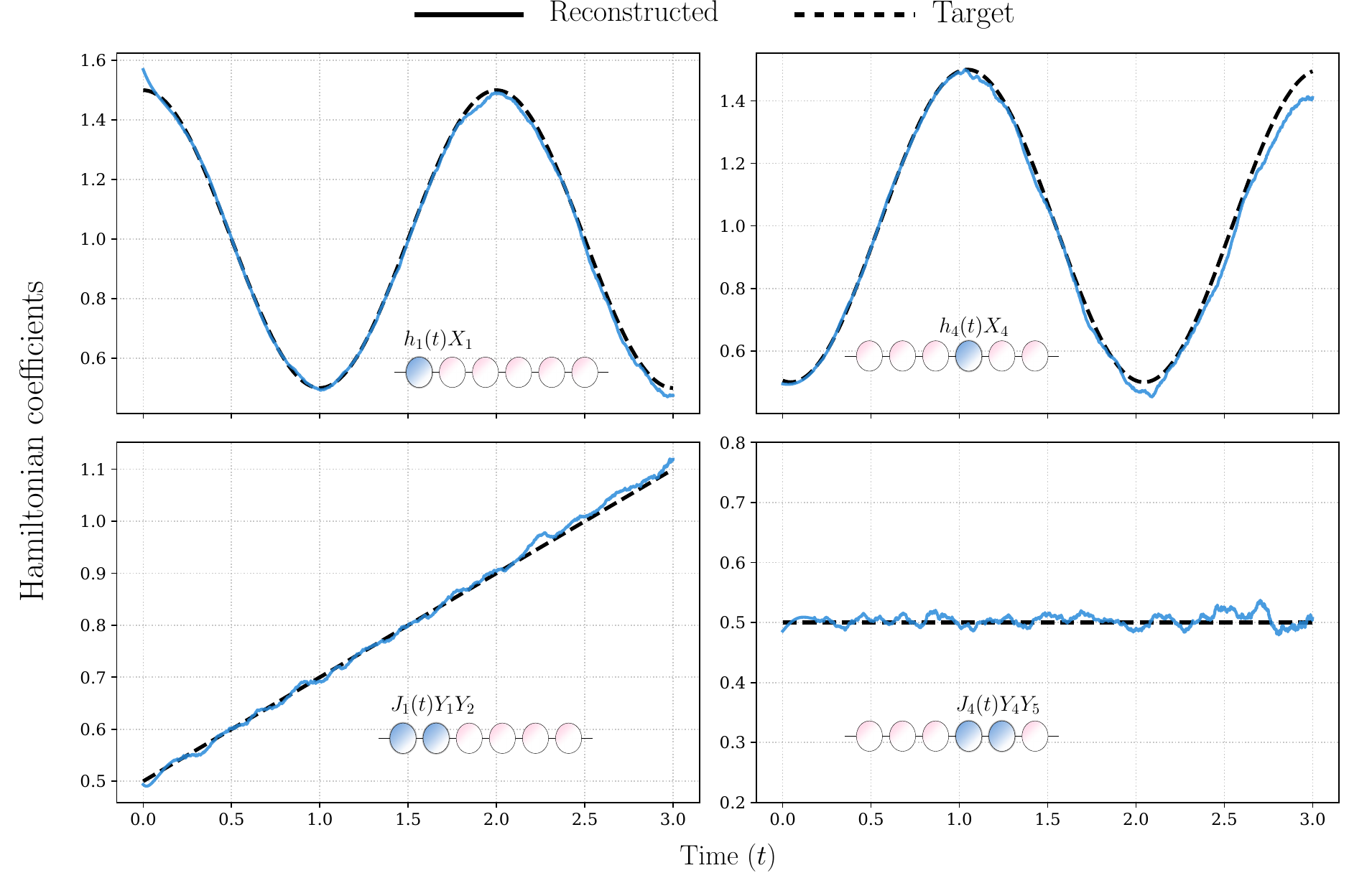}
    \caption{\justifying
    Time-dependent Hamiltonian reconstruction from weak-measurement signals. Dashed black curves show the target coefficients, while solid colored curves show the reconstructed trajectories. The panels display representative local fields $h_j(t)$ and nearest-neighbor couplings $J_j(t)$. The protocol tracks the coefficient trajectories over the reconstruction window, with larger fluctuations appearing for terms whose local inverse problems are more sensitive to noise and conditioning effects.
    }
    \label{fig:coefficient_reconstruction}
\end{figure*}

Figure~\ref{fig:coefficient_reconstruction} compares the reconstructed coefficients with the target trajectories. Out of the $2n-1 = 11$ coefficients in the $n = 6$ chain, we display four representative examples, including both local-field and coupling terms. The complete reconstruction of all $11$ coefficients is shown in Appendix~\ref{app:full_reconstruction}. This restriction is made solely for visual clarity: the protocol reconstructs the full coefficient vector, and the plotted terms are not selected because of any intrinsic limitation. The reconstructed trajectories closely follow the target ones over the full reconstruction window.  The coupling sector shows larger high-frequency fluctuations, consistent with the stronger sensitivity of those terms to finite-difference noise and local matrix conditioning. The largest deviations, of about $0.1$ for the local fields $h_4$ and $h_6$ near the end of the window, are instead deterministic: they persist when the simulation is repeated without signal noise and come from the accumulated error of the Euler state update (Sec.~\ref{sec:performance}).

\begin{figure*}[t]
    \centering
    \includegraphics[width=0.8\linewidth]{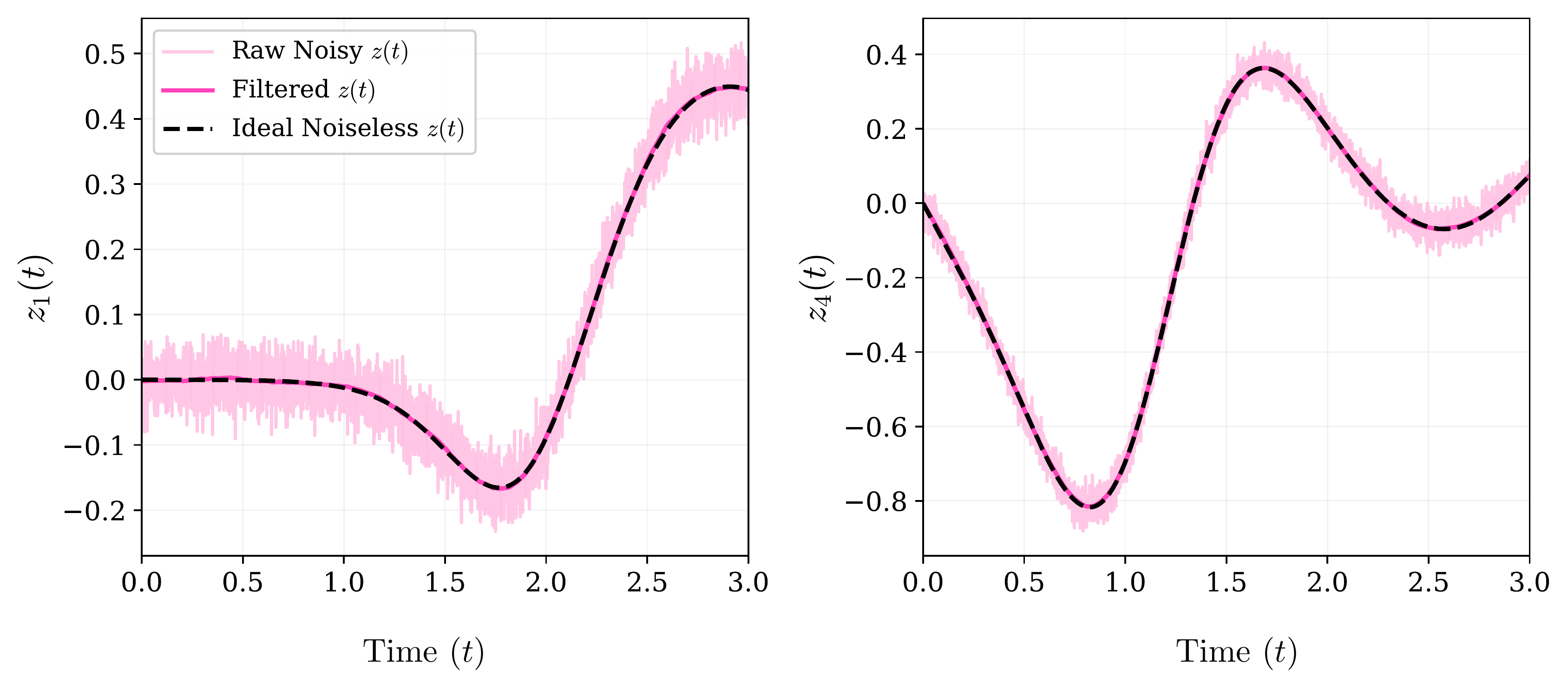}
    \caption{\justifying
    Preprocessing of averaged weak-measurement signals prior to reconstruction. The raw signals contain visible finite-sampling fluctuations, while the Savitzky--Golay filtered signals closely track the ideal noiseless trajectories. This filtering step stabilizes the finite-difference estimates used in the inversion protocol.}
    \label{fig:filtering_records}
\end{figure*}

Before applying the inverse step, the averaged weak-measurement signals are filtered. This preprocessing is necessary because the reconstruction uses finite differences of the measured signals, so high-frequency noise in $\hat z_j^{(s)}(t)$ is amplified by the factor $1/\Delta t$. Throughout the numerical analysis, we use a Savitzky--Golay filter to smooth the averaged signals while preserving the low-frequency dynamical features relevant for Hamiltonian reconstruction. Figure~\ref{fig:filtering_records} illustrates this preprocessing for two representative qubits: the filtered signal closely follows the ideal weak-measurement signal while significantly suppressing the finite-sampling fluctuations present in the raw data. Since Savitzky--Golay differentiation is linear in the signals, it is an instance of the derivative estimator in Eq.~\eqref{eq:finite_difference_record}, and the guarantees of Sec.~\ref{sec:performance} apply with the corresponding weights and constants $B$ and $q$. The filter also reduces the statistical cost, since smoothing lowers $\|w\|_2$ and hence $\kappa_w$, whose Gaussian part dominates for $\Gamma_d\Delta t\ll1$.

\textit{Probe ensemble dependence.}
\begin{figure*}[t!]
    \includegraphics[width=0.95\linewidth]{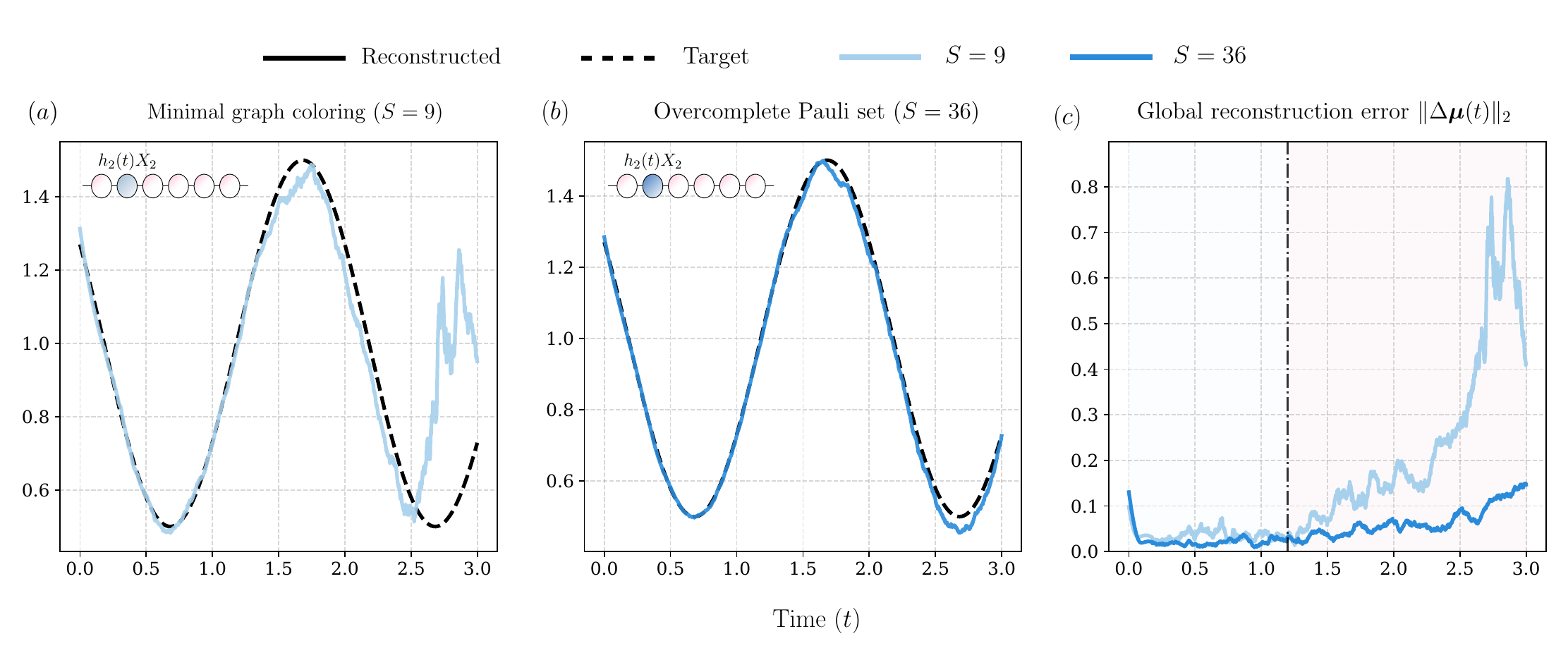}
    \caption{\justifying
    Probe-ensemble dependence of the reconstruction. 
    (a) Reconstruction of a representative local field $h_2(t)$ using the minimal graph-coloring ensemble with $S=9$ probe configurations. 
    (b) Reconstruction of the same coefficient using the overcomplete Pauli-eigenstate ensemble with $S=36$ probe configurations. 
    In both panels, the solid curve is the reconstructed coefficient and the dashed black curve is the target trajectory. 
    (c) Global reconstruction error $\|\Delta\boldsymbol{\mu}(t)\|_2$ for both ensembles. 
    The vertical dashed line at $t=1.2$ separates the short-time region, where both ensembles perform similarly, from the region where the minimal ensemble loses stability. The initial peak of the error for $t\lesssim0.2$ is a boundary effect of the Savitzky--Golay filter, which estimates the derivative using a one-sided window near $t=0$. 
    The minimal ensemble becomes unstable once the coherent evolution of the probe states makes some local inverse problems poorly conditioned. In contrast, the overcomplete ensemble remains stable over a longer time window.
    }
    \label{fig:probe_ensemble_comparison}
\end{figure*}
We next compare the minimal graph-coloring ensemble with the overcomplete Pauli-eigenstate ensemble, providing a numerical test of the preparation strategy introduced in Sec.~\ref{sec:probe_configurations}. For the TFIM-like chain, the graph-coloring construction gives a minimal separable ensemble with $S=9$ global probe configurations, while the overcomplete ensemble uses $S=36$ Pauli-product probes. The minimal ensemble is the most economical construction within the coloring strategy, since it activates the local inverse problems with the smallest number of preparations. This economy comes at the cost of a highly structured, non-uniform probe pattern, in which different qubits play different roles depending on the color class.

Figure~\ref{fig:probe_ensemble_comparison} shows the resulting trade-off. Panels~(a) and~(b) compare the reconstruction of the same representative local field $h_2(t)$ using the minimal and overcomplete ensembles. At short times, both ensembles track the target trajectory, confirming that the minimal graph-coloring probes contain enough information to initialize the local inversions. At later times, the difference becomes clear. The minimal reconstruction loses stability, while the overcomplete reconstruction remains close to the target coefficient over the full window.

The global error in Fig.~\ref{fig:probe_ensemble_comparison}(c) makes this behavior explicit. In the short-time region, both probe choices give comparable reconstruction errors. After the vertical dashed line, the error of the minimal ensemble grows steadily, exceeding $0.1$ at $t\approx1.5$ and rising sharply for $t\gtrsim2.5$, whereas the error of the overcomplete ensemble remains below $0.15$ over the whole window. This is not a contradiction with the graph-coloring construction. The construction guarantees that the local problems are identifiable at the initial time, but the reconstruction matrices at later times are built from the propagated probe states, which evolve coherently under the Hamiltonian. Since the minimal ensemble uses an asymmetric set of probes, this evolution can make some local reconstruction matrices poorly conditioned, which amplifies the statistical noise and the errors accumulated in the state propagation. Along the exact probe trajectories of Fig.~\ref{fig:probe_ensemble_comparison}, the smallest singular value of the local reconstruction matrices decreases from about $2.8$ at $t=0$ to about $0.3$ at $t=3$ for the minimal ensemble, while it remains between $1.2$ and $2.6$ for the overcomplete ensemble. The overcomplete ensemble is more robust because it samples the local operator space more symmetrically, spreading information across many probe directions rather than concentrating it in the minimal set.

This comparison highlights the practical trade-off in the probe preparation. The minimal graph-coloring ensemble is cheap and accurate over short reconstruction windows. For longer windows, additional probe configurations become advantageous. The overcomplete ensemble increases the preparation cost, but provides a more symmetric and stable reconstruction under finite-sampling fluctuations and propagation errors.
\begin{figure*}[t!]
    \includegraphics[width=0.95\linewidth]{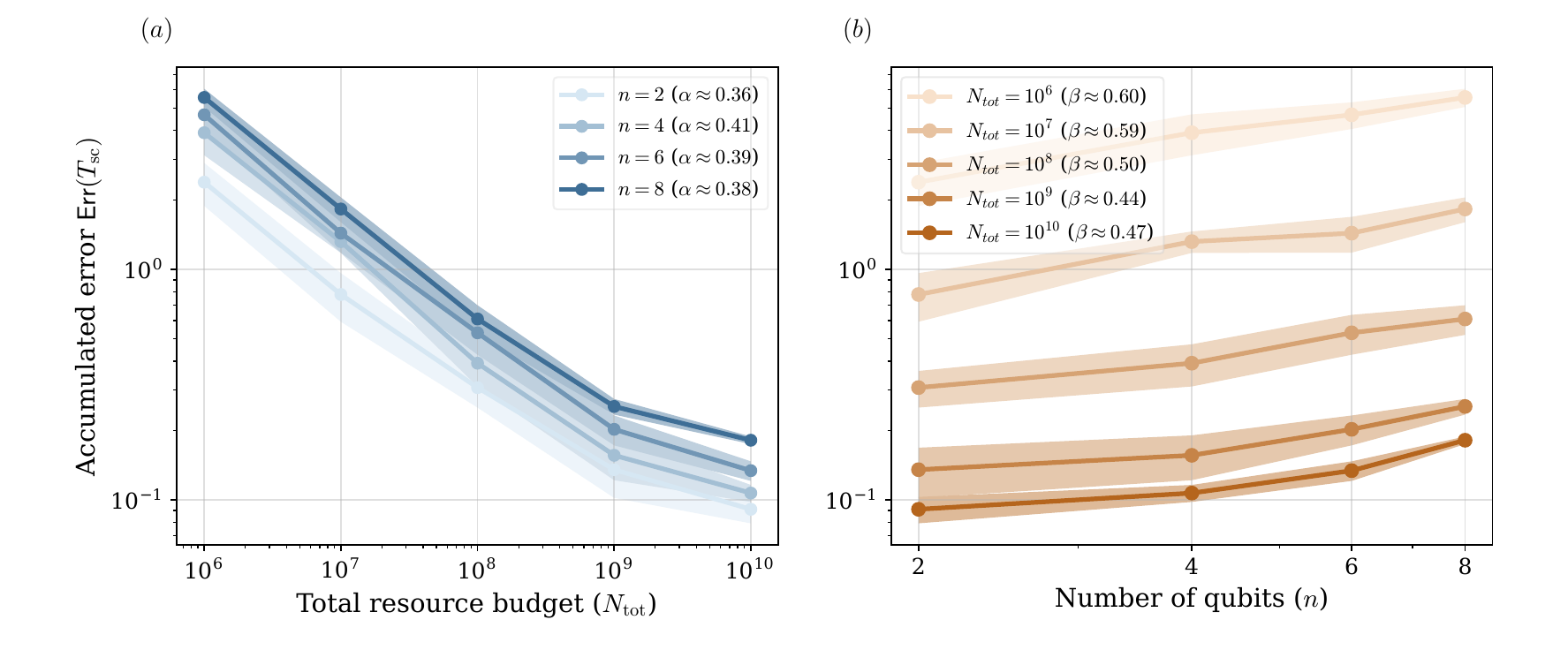}
    \caption{\justifying
    Sample-complexity and system-size scaling of the reconstruction protocol. (a) Accumulated global coefficient error $\Err(T_{\mathrm{sc}})$, restricted to the short-time window $T_{\mathrm{sc}}=0.3$, as a function of the total trajectory budget $N_{\mathrm{tot}}=SN_s$ for chains with $n=2,4,6,8$ qubits. (b) The same error plotted as a function of the system size $n$ for fixed values of $N_{\mathrm{tot}}$. The probe ensemble consists of the minimal graph-coloring set supplemented by $11$ randomly sampled Pauli-product probes, giving $S=20$. Each point is averaged over $10$ independent realizations of the measurement signals, and the shaded regions indicate one standard deviation over these realizations. Panel (a) shows a power-law decrease $\Err\propto N_{\mathrm{tot}}^{-\alpha}$, with $\alpha\approx0.4$ when all budgets are fitted and $\alpha\approx0.45$--$0.50$ over the statistics-dominated budgets $N_{\mathrm{tot}}\leq10^8$, followed by a saturation due to the deterministic bias from derivative estimation, integration, and state propagation. Panel (b) shows the corresponding system-size scaling $\Err\propto n^\beta$, which remains sublinear over the range considered and is consistent with the global scaling predicted by the coefficient-error bound.
    }
    \label{fig:sample_complexity_scaling}
\end{figure*}

\textit{Sample-complexity scaling.}
We evaluate the protocol's sampling performance by analyzing how the reconstruction error changes with the total number of trajectories used to estimate the weak-measurement signals. To isolate this effect, we restrict the comparison to the short-time window $T_{\mathrm{sc}}=0.3$, where the deterministic propagation bias remains controlled. We also fix the probe ensemble across all system sizes, using the minimal graph-coloring set supplemented by $11$ randomly sampled Pauli-product probes. The effective number of probe configurations is therefore $S=20$, and the resource plotted in Fig.~\ref{fig:sample_complexity_scaling} is the total trajectory budget $N_{\mathrm{tot}}=SN_s$. For each value of $N_{\mathrm{tot}}$, the reconstruction is repeated over $10$ independent realizations of the measurement signals, and the shaded regions indicate one standard deviation over these realizations.

The reconstruction quality is quantified by the accumulated global coefficient error
\begin{equation}
    \Err(T_{\mathrm{sc}})
    =
    \left(
    \sum_{k:\,t_k< T_{\rm sc}}
    \left\|
    \hat{\boldsymbol{\mu}}(t_k)
    -
    \boldsymbol{\mu}(t_k)
    \right\|_2^2
    \right)^{1/2}.
    \label{eq:numerical_global_error}
\end{equation}
This metric measures the reconstruction error of the full coefficient trajectory over a short window, where the deterministic bias remains small. Since $\Err$ is a global coefficient-vector error, larger systems naturally accumulate contributions from more local inverse problems. The vertical separation between system sizes in Fig.~\ref{fig:sample_complexity_scaling}(a) is therefore consistent with the global $\sqrt n$ dependence appearing in the coefficient-error bound. This is tested directly in Fig.~\ref{fig:sample_complexity_scaling}(b), where the same error is plotted as a function of $n$ for fixed trajectory budgets. The fitted exponents remain sublinear, $\beta\simeq0.44$--$0.60$. For the statistics-dominated budgets $N_{\mathrm{tot}}\leq10^8$, $\beta\simeq0.50$--$0.60$, close to the $\sqrt n$ dependence of the bound. At the largest budgets the error is dominated by the deterministic bias floor discussed below, whose own size dependence gives $\beta\approx0.48$, so the exponents there reflect the bias rather than the statistical term.

Figure~\ref{fig:sample_complexity_scaling} shows $\Err$ as a function of $N_{\mathrm{tot}}$ for chains with $n=2,4,6,8$ qubits. For small and intermediate budgets, increasing $N_{\mathrm{tot}}$ reduces the finite-sampling fluctuations in the weak-measurement signals, and the error follows an approximate power law $\Err\propto N_{\mathrm{tot}}^{-\alpha}$ with fitted exponents $\alpha\simeq0.36$--$0.41$ when all budgets are included. These fits contain the onset of the bias floor at the largest budgets. Restricted to the statistics-dominated budgets $N_{\mathrm{tot}}=10^6$--$10^8$, they give $\alpha\simeq0.45$--$0.50$, consistent with the $N_{\mathrm{tot}}^{-1/2}$ scaling of the sampling term in Theorem~\ref{thm:sample_complexity}. Evaluating the statistical term of Eq.~\eqref{eq:coefficient_error_bound} with the parameters of these simulations, with $\kappa_w$ computed from the Savitzky--Golay weights and the smallest singular value of the local reconstruction matrices computed along the probe trajectories, gives values that exceed the observed errors by a factor of $10$ to $30$. This is the expected overhead of a concentration bound combined with a union bound over all monitored sites and probe configurations.

 The crossover at large trajectory budgets is the clearest numerical signature of the time-step trade-off discussed in Sec.~\ref{sec:performance}. Once the sampling noise has been sufficiently suppressed, increasing $N_s$ no longer improves the reconstruction at the same rate. The curves approach a bias floor, which is the numerical manifestation of the deterministic contribution in Eq.~\eqref{eq:coefficient_error_bound}. The observed floor lies orders of magnitude below that contribution, which is dominated by the worst-case factor $e^{\gamma\bar Lt_k}$ discussed in Sec.~\ref{sec:performance}. This contribution is intrinsic to the iterative structure of the protocol. At each time step, the reconstructed Hamiltonian is used to propagate the probe states and build the next reconstruction matrices, so Euler discretization errors and previous coefficient errors are fed back into later inversions. Additional trajectories reduce the noise in the measured signals, but they cannot remove this accumulated state-update error. The Savitzky--Golay filter, whose window is fixed in time, adds a further bias to the derivative estimates that does not decrease with $N_s$ either. The log-log plot therefore separates the two mechanisms predicted by the theory: a sampling-limited regime, where more trajectories improve the reconstruction, and a propagation-limited regime, where the accuracy is controlled by the integration strategy, the reconstruction time step, and the stability of the dynamical update.

\section{Conclusion and outlook}\label{sec:conclusion}

In this work, we present a rigorous protocol for time-dependent HL based on continuous weak measurements. The protocol exploits the sparsity of the Hamiltonian to reduce the global reconstruction to local inverse problems, whose size is controlled by the interaction degree rather than by the full system size. These local problems are built directly from time-resolved weak-measurement signals and can be solved in parallel once suitable probe states are prepared. We show that separable probes are sufficient to initialize the local inversions and provide a systematic graph-coloring construction to embed them into global product-state preparations. As a result, the number of required probe configurations is governed by the local connectivity of the active-neighborhood graph, so the protocol extends to larger devices without increasing the number of state preparations, and repeated local structures allow the same probe pattern to be reused across the system.

We bound the reconstruction error at each time step as a function of the number of trajectories and the time step used in the iterative state update. The resulting bound separates the statistical error associated with finite weak-measurement signals from the deterministic bias arising from the derivative approximation and state propagation. This separation reveals the central trade-off of the protocol: decreasing $\Delta t$ reduces the bias of the derivative estimator and of the integration scheme but amplifies the finite-difference noise, whereas increasing $\Delta t$ reduces sampling noise but increases these biases. The bounds hold for any linear derivative estimator and any explicit integration scheme of order $p$, with forward differences and the Euler update as the simplest instance. The corresponding sample-complexity theorem gives sufficient conditions on the number of trajectories per probe configuration required to achieve a prescribed global reconstruction precision.

We validate the protocol through numerical simulations of time-dependent TFIM-like Hamiltonians in chains of up to $n=8$ qubits. The simulations show accurate reconstruction of the Hamiltonian coefficient trajectories from weak-measurement signals and confirm the role of the probe ensemble in the stability of the method. The minimal graph-coloring ensemble is sufficient over short time windows, while overcomplete separable Pauli-product probes provide a more symmetric sampling of the local operator space and remain robust for longer reconstruction windows. The numerical scaling analysis also reproduces the two regimes predicted by the theory. For limited trajectory budgets, the error decreases with the number of samples, while at large budgets the improvement saturates at a deterministic bias floor set by the derivative estimation, the Euler integration, and the accumulated state-update error. These results show that the protocol is experimentally friendly and structurally scalable at the level of the local inversion and probe-preparation steps.

Our protocol is a rigorous tool for characterizing and calibrating many-body quantum devices and is directly applicable to current quantum hardware. It only requires separable state preparations and local continuous readout, with no additional operations on the qubits during the evolution. It can be implemented natively on any platform that operates with dispersive readout or continuous monitoring, such as superconducting circuits, and on digital processors through ancilla-mediated weak measurements~\cite{franceschetto2026measurementenabledonlinequantumprocessing}. Reconstructing the full trajectory $H(t)$ can reveal distortions of the control pulses, such as amplitude and phase errors, as well as microwave crosstalk and unwanted exchange-type couplings between qubits, providing feedback for pulse calibration and guiding the design of improved devices. Terms that commute with the monitored observables, such as qubit detunings or $ZZ$ couplings, are not accessible at first order and require additional measurement bases \cite{PRXQuantum.4.040324}. 

The present implementation still leaves room for improving the deterministic part of the error. In our numerical simulations, the Euler update is used to propagate the reconstructed probe states, which contributes to the systematic bias observed in the numerical scaling analysis. This bias is not reduced by increasing the number of trajectories. Since the guarantees of Sec.~\ref{sec:performance} hold for any explicit scheme of order $p$, a natural improvement is to replace the first-order update by higher-order or adaptive integration methods, for which the integration bias scales as $\mathcal O(\Delta t^{p})$. Such integrators could lower the deterministic bias floor without changing the local inverse-problem structure of the protocol. 

A second direction concerns the classical simulation of the probe-state dynamics. The local inversions and probe preparation scale with the local interaction structure, but the numerical propagation used in this work still acts on the full density operator. Scaling the protocol to larger many-body systems therefore requires compressed dynamical representations. Tensor-network methods provide the natural route, using matrix-product-state or matrix-product-operator representations together with time-evolution algorithms such as TEBD or TDVP, and practical implementations such as TeNPy~\cite{Vidal2004TEBD, Haegeman2016TDVP, Hauschild2018TeNPy}. This would preserve the locality advantage of the reconstruction while avoiding explicit propagation in the full Hilbert space whenever the relevant trajectories admit an efficient tensor-network representation.

The time dependence of the coefficients could also be exploited. If the coefficients admit a compact expansion in a known basis, as in the HL protocol proposed in Ref.~\cite{StilckFranca2025LocalTimeDependent} and in the tomography of parametrized quantum states~\cite{Schreiber2025ParametrizedTomography}, the equations at all time steps can be combined into a single regression, which averages the noise over the whole window and can be solved by compressed-sensing techniques~\cite{Donoho2006CompressedSensing,Gross2010CompressedSensingTomography,Shabani2011CompressiveHamiltonian}. Nevertheless, the resulting problem is nonlinear because the system of equations to be solved depends on the unknown state trajectory. This approach could lead to tighter or even a priori error bounds for short evolution times, and potentially provide a quadratic advantage over evolve-and-measure strategies when combined with quantum control, as shown for time-independent Hamiltonians~\cite{Dutkiewicz2024QuantumControl}.

The probe design can also be improved. In this work, we restrict ourselves to separable probes in order to keep the protocol experimentally friendly. However, this is not necessarily optimal over all possible quantum preparations. Entangled probes could reduce the number of configurations or improve the conditioning of the local reconstruction matrices. A complementary strategy is to keep product-state accessibility but choose the probes in a more uniform and structured way. Mutually unbiased bases, projective and unitary designs, stabilizer states, and Clifford-generated ensembles provide natural candidates for such overcomplete probe families~\cite{Klappenecker2005MUBDesigns,Dankert2009UnitaryDesigns,Gross2007UnitaryDesigns,Zhu2015MUBCliffordDesigns,Lami2025CliffordMPSDesigns}. The numerical comparison between the minimal and overcomplete ensembles already suggests that a more symmetric coverage of the local operator space improves the conditioning of the local reconstruction matrices at later times.

Finally, the TFIM-like chains studied here are only a first benchmark. The protocol itself is not restricted to one-dimensional geometries. It applies to any sparse Hamiltonian whose active-neighborhood graph has bounded degree, including higher-dimensional qubit arrays, heavy-hex layouts, neutral-atom arrays, and other sparse coupling graphs relevant to current quantum devices~\cite{Arute2019QuantumSupremacy,Chamberland2020HeavyHex,Browaeys2020OpticalTweezers,Ebadi2021QuantumPhases}. Extending the numerical analysis to these geometries is a natural next step. Beyond HL, the same weak-measurement philosophy could also be developed toward time-dependent Lindbladian learning, where the dissipative part of the generator is unknown. We leave this open-system extension for future work.

\section*{Code availability statement}
All code for the simulations is available at \cite{git_tuto}.

\begin{acknowledgments} 
This project has received funding from  the Government of Spain (Severo Ochoa CEX2019-000910-S, FUNQIP and QEC4QEA PCI2025-163167), Fundació Cellex, Fundació Mir-Puig, Generalitat de Catalunya (CERCA program),  European Union (NextGenerationEU PRTR-C17.I1,
PASQuanS2.1, 101113690 and QEC4QEA, 101194322), 
and the AXA Chair in Quantum Information Science. 
G.F. acknowledges support from a “la
Caixa” Foundation (ID 100010434) fellowship. The fellowship code is LCF/BQ/DI23/11990070.
\end{acknowledgments}
\bibliography{bib}

\newpage
\onecolumngrid
\appendix
\section{Details of the reconstruction-error bound}
\label{app:error_bounds}
This appendix provides the technical derivation of Proposition~\ref{prop:coefficient_error_bound} and Theorem~\ref{thm:sample_complexity}. We first collect the norm and concentration tools used throughout the proof, then derive the data-vector concentration bound and the perturbation bound for the reconstruction matrices separately. Finally, we combine the local estimates into the global coefficient-error bound.

\subsection{Norms and concentration tools}
\label{app:norms_concentration_tools}

We first recall the norm conventions used in the derivation.

\begin{defn}[Induced matrix norm]
\label{def:induced_matrix_norm}
Let $\|\cdot\|$ be a norm on a vector space. The matrix norm induced by $\|\cdot\|$ is defined as
\begin{equation}
    \|A\|:=\sup_{x\neq 0}\frac{\|Ax\|}{\|x\|}=\sup_{\|x\|=1}\|Ax\|.
\end{equation}
In particular, the norm induced by the Euclidean norm is the spectral norm, which coincides with the Schatten $\infty$-norm and is denoted by $\|A\|_\infty$.
\end{defn}

\begin{defn}[Schatten norms]
\label{def:schatten_norm}
Let $A$ be a matrix with nonzero singular values $\{\sigma_i(A)\}_{i=1}^{\operatorname{rank}(A)}$. For $1\leq p<\infty$, the Schatten $p$-norm is
\begin{equation}
    \|A\|_{p}=\left(\sum_{i}\sigma_i(A)^p\right)^{1/p}.
\end{equation}
For $p=\infty$, $\|A\|_{\infty}=\max_i\sigma_i(A)$. Moreover, for $1\leq p\leq q\leq\infty$,
\begin{equation}
    \|A\|_q\leq\|A\|_p\leq \operatorname{rank}(A)^{1/p-1/q}\|A\|_q.
    \label{eq:schatten_norm_inequality}
\end{equation}
\end{defn}

\begin{thm}[Hölder's inequality for Schatten norms]
\label{thm:holder_schatten}
Let $A$ and $B$ be matrices, and let $p,q,r\in[1,\infty]$ satisfy $1/r=1/p+1/q$. Then
\begin{equation}
    \|AB\|_{r}\leq\|A\|_{p}\|B\|_{q}.
\end{equation}
In particular, for any observable $O$ and trace-class operator $X$,
\begin{equation}
    |\mathrm{Tr}(OX)|\leq\|O\|_{\infty}\|X\|_{1}.
    \label{eq:holder_trace_bound}
\end{equation}
\end{thm}

\begin{defn}[Lipschitz continuity]
\label{def:lipschitz_continuity}
Let $(\mathcal X,\|\cdot\|_{\mathcal X})$ and $(\mathcal Y,\|\cdot\|_{\mathcal Y})$ be normed vector spaces. A map $F:\mathcal X\rightarrow\mathcal Y$ is Lipschitz continuous with constant $L\geq0$ if
\begin{equation}
    \|F(x)-F(y)\|_{\mathcal Y}\leq L\|x-y\|_{\mathcal X}
\end{equation}
for all $x,y\in\mathcal X$.
\end{defn}

\begin{thm}[Hoeffding's inequality]
\label{thm:hoeffding}
Let $X_1,\ldots,X_N$ be independent random variables with common mean, which are sub-Gaussian with parameter $s$, i.e., $\mathbb E\,e^{\lambda(X_k-\mathbb E[X_k])}\leq e^{\lambda^2s^2/2}$ for all $\lambda\in\mathbb R$. Then, for every $\epsilon>0$,
\begin{equation}
    \Pr\!\left(\left|\frac{1}{N}\sum_{k=1}^{N}X_k-\mathbb E[X_k]\right|\geq\epsilon\right)
    \leq 2\exp\!\left(-\frac{N\epsilon^2}{2s^2}\right).
\end{equation}
\end{thm}
A variable with $a\leq X\leq b$ is sub-Gaussian with $s=(b-a)/2$, which recovers the standard form of Hoeffding's inequality, and a Gaussian variable with standard deviation $s$ is sub-Gaussian with parameter $s$. If $X$ and $Y$ are sub-Gaussian with parameters $s_X$ and $s_Y$, then $X+Y$ is sub-Gaussian with parameter $s_X+s_Y$, even if they are dependent, by Hölder's inequality.

\subsection{Concentration of the finite-difference data vector}
\label{app:data_vector_concentration}

We derive the concentration bound for the empirical finite-difference vectors. Fix a monitored site $j$, a probe configuration $s\in[S]$, and a reconstruction time $t_k$. Since the same trajectories are used at all the times $t_{k+i}$ entering Eq.~\eqref{eq:finite_difference_record}, the empirical finite difference is itself an average over trajectories,
\begin{equation}
    d\hat\Sigma_j^{(s)}(t_k)=\frac{1}{N_s}\sum_{\nu=1}^{N_s}X_\nu,
    \qquad
    X_\nu=\frac{1}{\Delta t}\sum_{i\in I}w_i\,z_{j,\nu}^{(s)}(t_{k+i}),
    \label{eq:appendix_finite_difference_triangle}
\end{equation}
where $z_{j,\nu}^{(s)}(t_k)$ is the rescaled homodyne current of Eq.~\eqref{eq:homodyne_current} in trajectory $\nu$. It is the sum of a signal, given by $\langle Z_j\rangle_{\mathrm c}$, with values in $[-1,1]$, and independent Gaussian noise with variance $1/(2\Gamma_j\Delta t)$. Hence $X_\nu$ is the sum of a part bounded by $\|w\|_1/\Delta t$ and a Gaussian part with standard deviation $\|w\|_2/(\Delta t\sqrt{2\Gamma_j\Delta t})$, so it is sub-Gaussian with parameter at most $\kappa_w/\Delta t$. The variables $X_\nu$ are independent and satisfy $\mathbb E[X_\nu]=d\Sigma_j^{(s)}(t_k)$. Applying Theorem~\ref{thm:hoeffding} gives
\begin{equation}
    \Pr\!\left(\left|d\hat\Sigma_j^{(s)}(t_k)-d\Sigma_j^{(s)}(t_k)\right|\geq\epsilon\right)
    \leq
    2\exp\!\left(-\frac{N_s\Delta t^2\epsilon^2}{2\kappa_w^2}\right).
    \label{eq:hoeffding_single_record_appendix}
\end{equation}
For fixed $j$, we apply a union bound over the $S$ probe configurations and use
$\|x\|_2\leq \sqrt{S}\|x\|_\infty$ for $x\in\mathbb R^S$. Hence, with probability at least $1-\delta$,
\begin{equation}
    \left\|d\hat{\boldsymbol{\Sigma}}_j(t_k)-d\boldsymbol{\Sigma}_j(t_k)\right\|_2
    \leq
    \frac{\kappa_w}{\Delta t}
    \sqrt{\frac{2S}{N_s}\log\!\left(\frac{2S}{\delta}\right)}.
    \label{eq:appendix_local_data_bound}
\end{equation}

For the global reconstruction bound, we need concentration to hold simultaneously across all monitored sites. Applying the same argument to the global stacked vector $d\boldsymbol{\Sigma}(t_k)=\left(d\boldsymbol{\Sigma}_1(t_k),\ldots,d\boldsymbol{\Sigma}_n(t_k)\right)^T\in\mathbb R^{nS}$ gives, with probability at least $1-\delta$,
\begin{equation}
    \left\|d\hat{\boldsymbol{\Sigma}}(t_k)-d\boldsymbol{\Sigma}(t_k)\right\|_2
    \leq
    \frac{\kappa_w}{\Delta t}
    \sqrt{\frac{2nS}{N_s}\log\!\left(\frac{2Sn}{\delta}\right)}.
    \label{eq:appendix_global_data_bound}
\end{equation}

\subsection{Perturbation of the reconstruction matrices}
\label{app:matrix_error_bound}

We now derive the deterministic matrix-perturbation bound used in Proposition~\ref{prop:coefficient_error_bound}. Fix a monitored site $j$ and a reconstruction time $t_k$. We suppress the explicit time dependence and write $\Delta\matM_j=\hat{\matM}_j-\matM_j$, where $\matM_j=\matM_j^{\mathrm{loc}}(t_k)$, $\hat{\matM}_j=\hat{\matM}_j^{\mathrm{loc}}(t_k)$, and $\rho_s=\rho_s(t_k)$. The entries of $\Delta\matM_j$ are
\begin{equation}
    \left(\Delta\matM_j\right)_{sm}
    =
    -i\,\mathrm{Tr}\!\left([Z_j,W_m](\hat\rho_s-\rho_s)\right),
    \qquad W_m\in\mathcal N(j).
    \label{eq:matrix_perturbation_entries}
\end{equation}
By Hölder's inequality, Eq.~\eqref{eq:holder_trace_bound},
\begin{equation}
    \left|\left(\Delta\matM_j\right)_{sm}\right|
    \leq
    \left\|[Z_j,W_m]\right\|_{\infty}\left\|\hat\rho_s-\rho_s\right\|_1
    \leq
    2\left\|\hat\rho_s-\rho_s\right\|_1,
\end{equation}
where we used $\|[Z_j,W_m]\|_{\infty}\leq2$ for Pauli strings.

Introduce the uniform state-propagation error
\begin{equation}
    \|\hat\rho-\rho\|_1=\max_{s\in[S]}\|\hat\rho_s-\rho_s\|_1.
\end{equation}
Using $\|\Delta\matM_j\|_{\infty}\leq\|\Delta\matM_j\|_2$ and $|\mathcal N(j)|\leq\mathcal D+1$, we obtain
\begin{equation}
    \|\Delta\matM_j\|_{\infty}
    \leq
    \|\Delta\matM_j\|_2
    \leq
    2\sqrt{S(\mathcal D+1)}\,\|\hat\rho-\rho\|_1.
    \label{eq:DM_bound}
\end{equation}

It remains to bound $\|\hat\rho-\rho\|_1$. In this part we denote the reconstruction time by $t_N=N\Delta t$ and use $k$ as the running time index. The true and reconstructed weak-measurement generators are
\begin{equation}
    \mathcal L_t(\rho)=-i[H(t),\rho]+\sum_{j=1}^{n}\frac{\Gamma_j}{2}D[Z_j]\rho,
    \qquad
    \hat{\mathcal L}_t(\rho)=-i[\hat H(t),\rho]+\sum_{j=1}^{n}\frac{\Gamma_j}{2}D[Z_j]\rho.
    \label{eq:true_and_reconstructed_generators}
\end{equation}
Thus, they differ only through the Hamiltonian term, and by Hölder's inequality, for any operator $X$,
\begin{equation}
    \left\|\left(\hat{\mathcal L}_{t}-\mathcal L_{t}\right)(X)\right\|_1
    =
    \left\|-i[\Delta H(t),X]\right\|_1
    \leq
    2\|\Delta H(t)\|_\infty\|X\|_1,
    \label{eq:liouvillian_difference_bound}
\end{equation}
where $\Delta H(t)=\hat H(t)-H(t)$. For the Pauli ansatz~\eqref{eq:hamiltonian_ansatz}, $\|\Delta H(t)\|_\infty\leq\|\hat{\boldsymbol{\mu}}(t)-\boldsymbol{\mu}(t)\|_1$, so this term carries no dimensional factor. In the sense of Definition~\ref{def:lipschitz_continuity}, assume that the true generator is Lipschitz continuous in trace norm, for all Hermitian operators $X$ and $Y$,
\begin{equation}
    \|\mathcal L_t(X)-\mathcal L_t(Y)\|_1\leq L\|X-Y\|_1.
    \label{eq:lipschitz_trace_norm}
\end{equation}
By Eq.~\eqref{eq:liouvillian_difference_bound}, the reconstructed generator $\hat{\mathcal L}_t$ is then Lipschitz continuous with constant $L+2\|\Delta H(t)\|_\infty$.

When the update from $t_k$ to $t_{k+1}$ is performed, $\hat H$ is only available at the reconstruction times $t_{k'}\leq t_k$, and the probe states only at the previous time steps. We therefore consider explicit causal schemes
\begin{equation}
    \rho_{k+1}=\rho_k+\Delta t\,\Psi_k[H](\rho_k,\ldots,\rho_{k-\ell+1}),
    \label{eq:causal_scheme}
\end{equation}
where the increment $\Psi_k[H]$ is a linear map built from the generator, with the Hamiltonian evaluated at the times $t_{k'}\leq t_k$ or extrapolated from them. For $\ell=1$, it is a one-step scheme. For $\ell>1$, the first $\ell-1$ steps use a starting procedure, and, as is standard for multistep methods~\cite{HairerNorsettWanner1993}, we assume that the starting values are accurate to the order of the scheme, so that Eq.~\eqref{eq:euler_local_truncation_bound} below also holds for these steps.

\begin{defn}[Integration scheme of order $p$]
\label{def:integrator}
The local truncation error of the scheme~\eqref{eq:causal_scheme} is the error made in a single step when the scheme is applied to the exact states $\rho(t_k),\ldots,\rho(t_{k-\ell+1})$, with the true Hamiltonian. The scheme has order $p$ if there is a constant $C\geq0$ such that, for all $k$ with $t_k<t_N$, its local truncation error satisfies
\begin{equation}
    \left\|\rho(t_k)+\Delta t\,\Psi_k[H]\big(\rho(t_k),\ldots,\rho(t_{k-\ell+1})\big)-\rho(t_{k+1})\right\|_1\leq C\Delta t^{p+1},
    \label{eq:euler_local_truncation_bound}
\end{equation}
where $\rho(t)$ is the exact trajectory generated by $\mathcal L_t$.
\end{defn}
The Euler update, $\Psi_k[H](X)=\mathcal L_{t_k}(X)$, has order $p=1$, with $C$ a uniform bound on the second-order remainder of the exact evolution over the reconstruction window. The exponential $e^{\Delta t\,\mathcal L_{t_k}}$ also has order one. Both schemes freeze $H$ during the step, so higher orders require information on $H$ at earlier reconstruction times. Runge--Kutta methods obtain it by extrapolating $\hat H$ to intermediate times, and Adams--Bashforth methods~\cite{HairerNorsettWanner1993} obtain it from the previous states, with $\ell=p$ and $\Psi_k[H](\rho_k,\ldots,\rho_{k-p+1})=\sum_{i=0}^{p-1}\beta_i\,\mathcal L_{t_{k-i}}(\rho_{k-i})$. Multistep schemes without the term $\rho_k$, which require the root condition for stability~\cite{HairerNorsettWanner1993}, and predictor--corrector schemes that use $\hat H(t_{k+1})$ are not considered here.

For a fixed probe trajectory, compare the exact continuous trajectory with the scheme driven by the reconstructed Hamiltonian,
\begin{align}
    \dot\rho(t) &= \mathcal L_t(\rho(t)), & \rho(0)&=\rho_0, \label{eq:exact_continuous_dynamics}\\
    \hat\rho_{k+1} &= \hat\rho_k+\Delta t\,\Psi_k[\hat H](\hat\rho_k,\ldots,\hat\rho_{k-\ell+1}), & \hat\rho_0&=\rho_0. \label{eq:reconstructed_euler_dynamics}
\end{align}
Let $e_k=\hat\rho_k-\rho(t_k)$ and $E_k=\max_{k'\leq k}\|e_{k'}\|_1$. Subtracting the exact trajectory from the scheme gives
\begin{equation}
\begin{aligned}
    e_{k+1}
    ={}&
    e_k
    +\Delta t\Big(\Psi_k[H](\hat\rho_k,\ldots)-\Psi_k[H](\rho(t_k),\ldots)\Big)
    \\
    &+\Delta t\Big(\Psi_k[\hat H]-\Psi_k[H]\Big)(\hat\rho_k,\ldots)
    +\tau_k,
\end{aligned}
    \label{eq:state_error_recursion}
\end{equation}
where $\tau_k=\rho(t_k)+\Delta t\,\Psi_k[H](\rho(t_k),\ldots,\rho(t_{k-\ell+1}))-\rho(t_{k+1})$ is the local truncation error of the scheme, bounded by Eq.~\eqref{eq:euler_local_truncation_bound}. The second term propagates the state error through the true dynamics, and the third is the learning error, coming from the reconstructed Hamiltonian.

To bound the second and third terms, we compare the increment $\Psi_k$ with a single evaluation of the generator. For a single evaluation, an error in the state is amplified at most by the Lipschitz constant $L$ of Eq.~\eqref{eq:lipschitz_trace_norm}, defined in the sense of Definition~\ref{def:lipschitz_continuity}, and an error in the Hamiltonian enters as $2\|\Delta H\|_\infty\|X\|_1$, by Eq.~\eqref{eq:liouvillian_difference_bound}. The increment combines several evaluations, weighted by the coefficients of the scheme, and can amplify both errors further. We quantify this amplification by the smallest constant $\gamma\geq1$ such that, for all $k$ with $t_k<t_N$ and all Hermitian operators $X_i,Y_i$,
\begin{equation}
\begin{aligned}
    \|\Psi_k[H](X)-\Psi_k[H](Y)\|_1&\leq\gamma L\max_i\|X_i-Y_i\|_1,\\
    \|\Psi_k[\hat H](X)-\Psi_k[H](X)\|_1&\leq2\gamma\max_{k'\leq k}\|\Delta H(t_{k'})\|_\infty\max_i\|X_i\|_1.
\end{aligned}
    \label{eq:gamma_definition}
\end{equation}
Thus $\gamma L$ is the Lipschitz constant of the increment, in the sense of Definition~\ref{def:lipschitz_continuity}, and $\gamma$ follows from Eqs.~\eqref{eq:liouvillian_difference_bound} and~\eqref{eq:lipschitz_trace_norm} and the absolute values of the coefficients of the scheme. For the Euler update, which uses a single evaluation, $\gamma=1$. Higher-order causal schemes use information on $H$ at several past times, combined with coefficients of alternating sign, which gives $\gamma>1$. For Adams--Bashforth methods, $\gamma=\sum_i|\beta_i|$, and for Runge--Kutta methods with extrapolation, $\gamma$ is set by the extrapolation coefficients. In both cases, $\gamma$ grows with $p$, so increasing the order reduces the local truncation error but amplifies error propagation.

The iterates $\hat\rho_k$ are Hermitian with unit trace, but they need not be positive semidefinite. We therefore write $\hat\rho_k=\rho(t_k)+e_k$ in the learning term, so that Eq.~\eqref{eq:gamma_definition} is applied with $\|\rho(t_k)\|_1=1$ and $\|e_k\|_1\leq E_k$. Since $E_k$ bounds the errors of all the states used by the scheme, Eqs.~\eqref{eq:euler_local_truncation_bound} and~\eqref{eq:gamma_definition} give
\begin{equation}
    E_{k+1}\leq(1+\gamma\bar L\Delta t)E_k+2\gamma\Delta t\,\Delta H_{\max}(t_{k+1})+C\Delta t^{p+1},
    \label{eq:state_error_gronwall}
\end{equation}
with $\bar L=L+2\Delta H_{\max}(t_N)$ and $\Delta H_{\max}(t_N)=\max_{0\leq k<N}\|\Delta H(t_k)\|_\infty$. Only coefficients reconstructed at earlier times enter, since $\hat\rho_N$ is obtained from $\hat H(t_0),\ldots,\hat H(t_{N-1})$. Iterating from $E_0=0$, summing the geometric series, and using $1+x\leq e^{x}$ and $\gamma\geq1$ gives
\begin{equation}
\begin{aligned}
    \|\hat\rho_N-\rho(t_N)\|_1\leq E_N
    &\leq
    \left(2\gamma\Delta t\,\Delta H_{\max}(t_N)+C\Delta t^{p+1}\right)\sum_{k'=0}^{N-1}(1+\gamma\bar L\Delta t)^{k'}
    \\
    &\leq
    \frac{e^{\gamma\bar Lt_N}-1}{\bar L}
    \left(2\Delta H_{\max}(t_N)+C\Delta t^{p}\right).
\end{aligned}
    \label{eq:state_propagation_error_bound}
\end{equation}
The two source terms are the learning error, accumulated from the reconstructed Hamiltonian, and the discretization error. The latter is of order $\Delta t^{p}$, one power less than the local truncation error, since $N=t_N/\Delta t$ local errors accumulate.
Identifying $t_N$ with the reconstruction time $t_k$ of Proposition~\ref{prop:coefficient_error_bound} and substituting Eq.~\eqref{eq:state_propagation_error_bound} into Eq.~\eqref{eq:DM_bound} yields
\begin{equation}
    \left\|\hat{\matM}_j^{\mathrm{loc}}(t_k)-\matM_j^{\mathrm{loc}}(t_k)\right\|_{\infty}
    \leq
    2\sqrt{S(\mathcal D+1)}
    \frac{e^{\gamma\bar Lt_k}-1}{\bar L}
    \left(2\Delta H_{\max}(t_k)+C\Delta t^{p}\right).
    \label{eq:appendix_matrix_error_final}
\end{equation}
Multiplying by $\|\boldsymbol{\mu}_{\mathcal N(j)}(t_k)\|_2\leq\mu_{\max}$ gives Eq.~\eqref{eq:matrix_error_bound}.

\subsection{Global reconstruction bound and proof of Theorem~\ref{thm:sample_complexity}}
\label{app:global_bound_sample_complexity}

This section analyzes the asymptotic scaling of the full protocol and justifies the global reconstruction bound used in the main text. Let $\|\Delta\boldsymbol{\mu}(t_k)\|$ denote the norm of the difference between the reconstructed and true coefficients of the global Hamiltonian, and $\|\Delta\boldsymbol{\mu}_j(t_k)\|$ the corresponding norm restricted to the local inverse problem at site $j$. The number of local inverse problems is $n$, one for each monitored site.

From the assembly of the global coefficient vector from the local estimates, we use
\begin{equation}
    \|\Delta\boldsymbol{\mu}(t_k)\|_2
    \leq
    \left(
    \sum_{j=1}^{n}
    \|\Delta\boldsymbol{\mu}_j(t_k)\|_2^2
    \right)^{1/2}.
    \label{eq:global_from_local_errors_l2}
\end{equation}
Let all local thresholds be equal to $\varepsilon_{\mathrm{loc}}$. Then
\begin{align}
    \Pr\!\left[
    \|\Delta\boldsymbol{\mu}(t_k)\|_2
    \geq
    \sqrt n\,\varepsilon_{\mathrm{loc}}
    \right]
    &\leq
    \Pr\!\left[
    \left(
    \sum_{j=1}^{n}
    \|\Delta\boldsymbol{\mu}_j(t_k)\|_2^2
    \right)^{1/2}
    \geq
    \sqrt n\,\varepsilon_{\mathrm{loc}}
    \right]
    \nonumber\\
    &\leq
    \sum_{j=1}^{n}
    \Pr\!\left[
    \|\Delta\boldsymbol{\mu}_j(t_k)\|_2
    \geq
    \varepsilon_{\mathrm{loc}}
    \right].
    \label{eq:global_union_bound_l2}
\end{align}
The first inequality follows from Eq.~\eqref{eq:global_from_local_errors_l2}, and the second from the union bound. Therefore, to guarantee a global failure probability at most $\delta$, it is sufficient to require each local failure probability to be at most $\delta/n$. To target
\begin{equation}
    \|\Delta\boldsymbol{\mu}(t_k)\|_2\leq\epsilon,
\end{equation}
we set
\begin{equation}
    \varepsilon_{\mathrm{loc}}=\frac{\epsilon}{\sqrt n}.
\end{equation}

Using the local reconstruction bound derived from Eqs.~\eqref{eq:appendix_local_data_bound}, \eqref{eq:derivative_bias}, and~\eqref{eq:matrix_error_bound}, this local requirement is achieved if
\begin{equation}
    \epsilon
    \geq
    \frac{\sqrt n}{\sigma_{\min}(\hat{\matM})}
    \left[
    \frac{\kappa_w}{\Delta t}
    \sqrt{
    \frac{2S}{N_s}
    \log\!\left(
    \frac{2Sn}{\delta}
    \right)
    }
    +
    B_{\mathrm{det}}(t_k,\Delta t)
    \right].
    \label{eq:local_threshold_condition}
\end{equation}
The logarithmic factor contains $n$ because the local concentration bound is evaluated with local failure probability $\delta/n$. Solving Eq.~\eqref{eq:local_threshold_condition} for $N_s$ yields
\begin{equation}
    N_s
    \geq
    \frac{2\kappa_w^2Sn}{\Delta t^2}
    \frac{
    \log\!\left(
    \frac{2Sn}{\delta}
    \right)
    }{
    \left(
    \epsilon\sigma_{\min}(\hat{\matM})
    -
    \sqrt n\,B_{\mathrm{det}}(t_k,\Delta t)
    \right)^2
    }.
    \label{eq:appendix_sample_complexity_final}
\end{equation}
This is the sample-complexity condition stated in Theorem~\ref{thm:sample_complexity}. In the regime where the deterministic bias is negligible compared with the target accuracy, Eq.~\eqref{eq:appendix_sample_complexity_final} reduces to
\begin{equation}
    N_s
    =
    \mathcal O\!\left(
    \frac{\kappa_w^2Sn}{\Delta t^2\epsilon^2\sigma_{\min}(\hat{\matM})^2}
    \log\!\left(
    \frac{2Sn}{\delta}
    \right)
    \right),
\end{equation}
which is the scaling summarized in the main text.

\section{Numerical implementation details}
\label{app:numerical_details}

All numerical experiments are performed on the TFIM-like nearest-neighbor chain of Eq.~\eqref{eq:numerical_tfim_hamiltonian} of the main text, with time-dependent coefficients $h_j(t)=1+\frac{1}{2}\cos(\pi t+j-1)$ and $J_j(t)=\frac{1}{2}+\frac{1}{5}\,t\,\mathbbm{1}_{\{j\ \mathrm{odd}\}}$, and fixed dephasing rate $\Gamma_d=10^{-3}$ (in the units of Sec.~\ref{sec:numerics}, $\hbar=1$ and $\omega_0$).

For each probe state, the ensemble-averaged weak-measurement signal is generated by extending the discrete weak-measurement formalism of Refs.~\cite{Mujal2023WeakProjective,backaction} to time-dependent Hamiltonians. At each time step, the state is propagated with the midpoint unitary $U=e^{-iH(t+\Delta t/2)\Delta t}$ and subsequently acted on by the effective measurement-induced dephasing map, implemented as the element-wise multiplication $\rho_{ab}\mapsto e^{-(g^2/2)\,d_H(a,b)}\rho_{ab}$ in the computational basis, where $d_H(a,b)$ is the number of qubits in which the basis states $a$ and $b$ differ, and $g=\sqrt{2\Gamma_d\Delta t}$ is the measurement strength. Finite sampling statistics are incorporated by adding Gaussian noise of standard deviation $\sigma=\sqrt{(g^2+1)/(g^2N_s)}$, reproducing the average over $N_s$ independent trajectories.

This discrete model follows from the model of Sec.~\ref{subsec:weak_measurements} as follows. The dephasing map is applied to each qubit and multiplies its coherences by $e^{-g^2/2}=e^{-\Gamma_d\Delta t}$, which is the exact evolution under the dissipative term of Eq.~\eqref{eq:wm_master_equation} over one time step. Alternating it with the midpoint unitary is a splitting of Eq.~\eqref{eq:wm_master_equation}, with error of order $\Gamma_d\|H\|\Delta t^2$ per step. The noise follows from Eq.~\eqref{eq:homodyne_current}. Rescaling by $1/\sqrt{2\Gamma_d}$, the current of a single trajectory is a signal bounded by one plus Gaussian noise of variance $1/(2\Gamma_d\Delta t)=1/g^2$. The average over $N_s$ trajectories therefore has variance at most $(1+1/g^2)/N_s=\sigma^2$, which is the value used in the simulations. Since the corresponding standard deviation per trajectory, $\sqrt{1+1/g^2}$, does not exceed the sub-Gaussian parameter $1+1/g$ used in Appendix~\ref{app:data_vector_concentration}, the simulated signals satisfy the assumptions of Proposition~\ref{prop:coefficient_error_bound}.

The dynamics is simulated up to $t=3.5$, providing a $0.5$-unit buffer beyond the reconstruction window $t_{\mathrm{rec}}=3.0$ to suppress boundary effects introduced by numerical differentiation. Time derivatives are estimated using a third-order Savitzky--Golay filter with a window covering approximately $0.35$ time units for the reconstructions of Figs.~\ref{fig:coefficient_reconstruction}--\ref{fig:probe_ensemble_comparison} and~\ref{fig:full}. Local inverse problems are solved through the Moore--Penrose pseudoinverse computed by singular-value decomposition with cutoff $5\times10^{-3}$, and coefficients reconstructed in overlapping active neighborhoods are averaged. Probe states are propagated with an explicit Euler step using the time step $\Delta t=\eta\,[2\log(2Sn)/(N_s(\mathcal D+1))]^{1/4}$, which is the scaling of Eq.~\eqref{eq:time_step} for $\bar p=1$ and constant $\kappa_w$, with empirical prefactor $\eta=0.1$ ($\Delta t=7.0\times10^{-4}$ for $S=36$ and $6.6\times10^{-4}$ for $S=9$), while Hermiticity and unit trace are enforced after each update to eliminate numerical drift.

Two separable probe ensembles are compared throughout. The minimal graph-coloring ensemble ($S = 9$) is constructed from the three-coloring of the active-neighborhood graph discussed in Sec.~\ref{sec:probe_configurations}. For each color class $c\in\{1,2,3\}$, three global product states are prepared by tiling a local pattern $(|\phi_L\rangle,|\phi_T\rangle,|\phi_R\rangle)$ across the chain following the periodic assignment $i\bmod 3$, where the entries denote the states of the left-neighbor, target, and right-neighbor qubit of each active neighborhood in that class. The three local patterns $(|x_+\rangle,|y_+\rangle,|x_+\rangle)$, $(|y_-\rangle,|x_+\rangle,|z_+\rangle)$, and $(|z_+\rangle,|x_+\rangle,|y_-\rangle)$ are chosen to isolate local-field, left-coupling, and right-coupling terms, respectively.  The overcomplete Pauli-eigenstate ensemble ($S = 36$) consists of global product states $\rho = \bigotimes_{j=1}^n|\alpha_j\rangle\langle\alpha_j|$ with each $|\alpha_j\rangle$ drawn independently and uniformly from $\{|x_\pm\rangle,|y_\pm\rangle,|z_\pm\rangle\}$.

For the sample-complexity analysis, systems with $n\in\{2,4,6,8\}$ qubits are simulated using a fixed hybrid ensemble of $S=20$ probe states, consisting of the $9$ graph-coloring probes supplemented by $11$ randomly sampled Pauli-product states generated with a fixed random seed. The total trajectory budget is uniformly distributed among probe configurations, with $N_s=N_{\rm tot}/S$, where $N_{\rm tot}\in\{10^6,10^7,10^8,10^9,10^{10}\}$ and $\Delta t=0.02$ is fixed. The signals are simulated up to $t=1.0$, and derivatives are estimated with a third-order Savitzky--Golay filter with a window of $21$ points, covering approximately $0.4$ time units. Performance is quantified through the accumulated reconstruction error $$\Err(T_{\rm sc})=\left(\sum_{k:\,t_k< T_{\rm sc}}\|\hat{\boldsymbol{\mu}}(t_k)-\boldsymbol{\mu}(t_k)\|_2^2\right)^{1/2}$$ over the conservative window $T_{\rm sc}=0.3$, after discarding the initial half-filter window to remove differentiation transients. The error is therefore evaluated for $0.2\leq t_k<T_{\rm sc}$, where each derivative estimate uses a complete, centered filter window. Each data point is obtained from $10$ independent realizations of the measurement signals, and the scaling exponents are extracted by linear regression in logarithmic scale.

\section{Full coefficient reconstruction}
\label{app:full_reconstruction}
\begin{figure}[h]
    \includegraphics[width=0.75\textwidth]{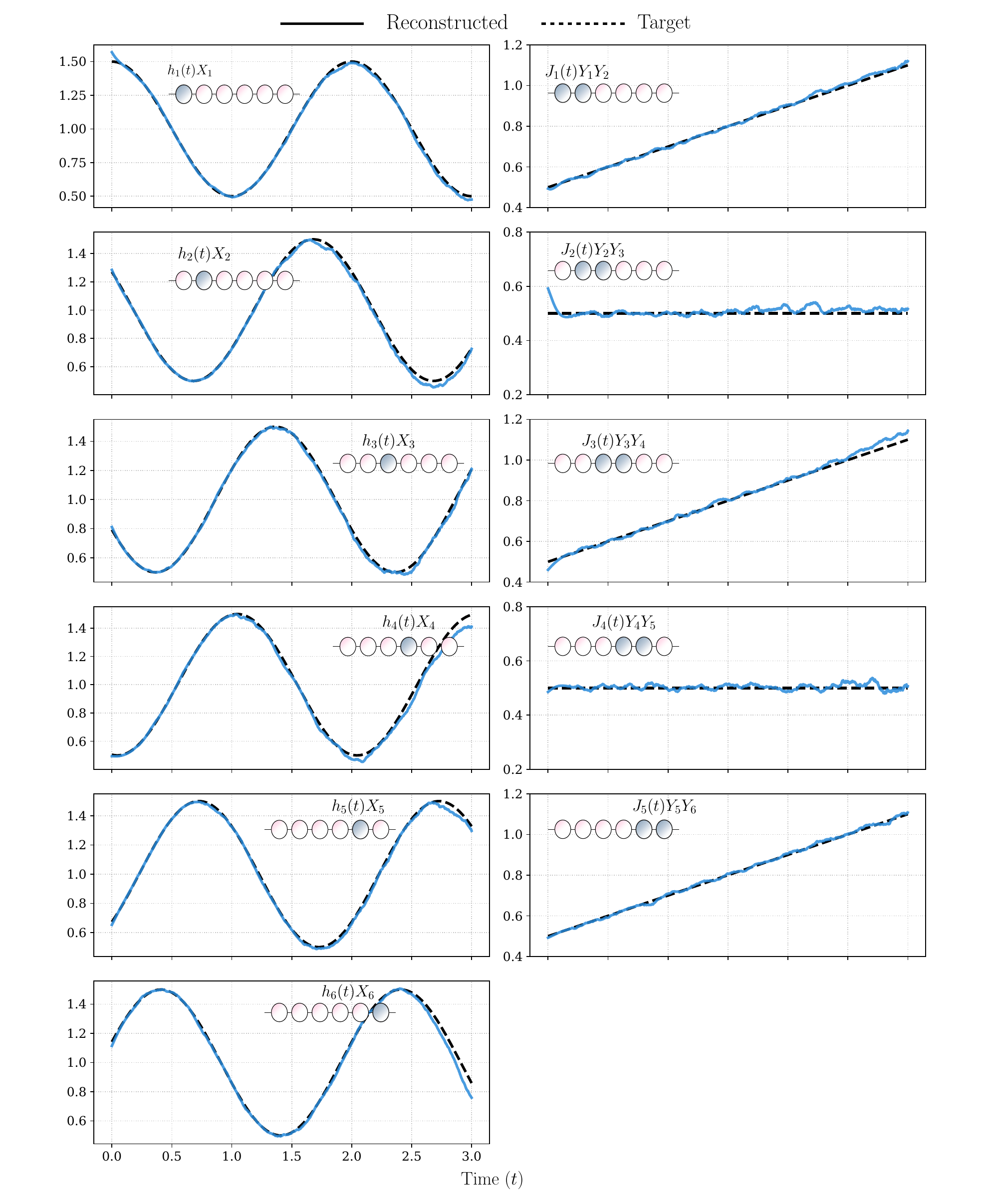}
    \caption{\justifying
    Full reconstruction of the $11$ Hamiltonian coefficients for the representative $n=6$ TFIM-like model. The left column shows the local-field coefficients, while the right column shows the nearest-neighbor coupling coefficients. In each panel, the reconstructed trajectory is compared against the target coefficient over the full reconstruction window. These results complement Fig.~\ref{fig:coefficient_reconstruction} of the main text, where only four representative coefficients were displayed for clarity.
    }\label{fig:full}
\end{figure}

We include here the complete reconstruction of the $2n-1=11$ Hamiltonian coefficients for the representative $n=6$ TFIM-like model considered in Sec.~\ref{sec:numerics}.
While the main text displays only four representative coefficients for clarity, Fig.~\ref{fig:full} shows the full set of reconstructed local-field and coupling trajectories.

\end{document}